\documentclass[twocolumn]{aastex631}

\newcommand{\Msun} {$\mathrm{M}_{\sun}$}

\begin{document}

\title{PHANGS-JWST First Results: Rapid Evolution of Star Formation in the Central Molecular Gas Ring of NGC\,1365}

\correspondingauthor{E. Schinnerer}
\email{schinner@mpia.de}

\suppressAffiliations
\author[0000-0002-3933-7677]{Eva Schinnerer}
\affiliation{Max Planck Institute for Astronomy, K\"onigstuhl 17, 69117, Germany}

\author[0000-0002-6155-7166]{Eric Emsellem}
\affiliation{European Southern Observatory, Karl-Schwarzschild-Stra{\ss}e 2, 85748 Garching, Germany}
\affiliation{Univ Lyon, Univ Lyon1, ENS de Lyon, CNRS, Centre de Recherche Astrophysique de Lyon UMR5574, F-69230 Saint-Genis-Laval France}

\author[0000-0001-9656-7682]{Jonathan~D. Henshaw}
\affiliation{Astrophysics Research Institute, Liverpool John Moores University, 146 Brownlow Hill, Liverpool L3 5RF, UK}
\affiliation{Max-Planck-Institut f\"ur Astronomie, K\"onigstuhl 17, D-69117 Heidelberg, Germany}

\author[0000-0001-9773-7479]{Daizhong Liu}
\affiliation{Max-Planck-Institut f\"ur Extraterrestrische Physik (MPE), Giessenbachstr. 1, D-85748 Garching, Germany}

\author[0000-0002-6118-4048]{Sharon E. Meidt}
\affiliation{Sterrenkundig Observatorium, Universiteit Gent, Krijgslaan 281 S9, B-9000 Gent, Belgium}

\author[0000-0002-0472-1011]{Miguel Querejeta}
\affiliation{Observatorio Astron\'{o}mico Nacional (IGN), C/Alfonso XII, 3, E-28014 Madrid, Spain}

\author[0000-0001-5073-2267]{Florent Renaud}
\affiliation{Department of Astronomy and Theoretical Physics, Lund Observatory, Box 43, SE-221 00 Lund, Sweden}

\author[0000-0001-6113-6241]{Mattia C.\ Sormani}
\affiliation{Universit\"{a}t Heidelberg, Zentrum f\"{u}r Astronomie, Institut f\"{u}r Theoretische Astrophysik, Albert-Ueberle-Stra{\ss}e 2, D-69120 Heidelberg, Germany}

\author[0000-0003-0378-4667]{Jiayi~Sun}
\affiliation{Department of Physics and Astronomy, McMaster University, 1280 Main Street West, Hamilton, ON L8S 4M1, Canada}
\affiliation{Canadian Institute for Theoretical Astrophysics (CITA), University of Toronto, 60 St George Street, Toronto, ON M5S 3H8, Canada}

\author[0000-0002-4755-118X]{Oleg V. Egorov}
\affiliation{Astronomisches Rechen-Institut, Zentrum f\"{u}r Astronomie der Universit\"{a}t Heidelberg, M\"{o}nchhofstra\ss e 12-14, D-69120 Heidelberg, Germany}

\author[0000-0003-3917-6460]{Kirsten L. Larson}
\affiliation{AURA for the European Space Agency (ESA), Space Telescope Science Institute, 3700 San Martin Drive, Baltimore, MD 21218, USA}

\author[0000-0002-2545-1700]{Adam~K. Leroy}
\affiliation{Department of Astronomy, The Ohio State University, 140 West 18th Avenue, Columbus, Ohio 43210, USA}

\author[0000-0002-5204-2259]{Erik Rosolowsky}
\affiliation{Department of Physics, University of Alberta, Edmonton, Alberta, T6G 2E1, Canada}

\author[0000-0002-4378-8534]{Karin M. Sandstrom}
\affiliation{Center for Astrophysics \& Space Sciences, University of California, San Diego, 9500 Gilman Drive, San Diego, CA 92093, USA}

\author[0000-0002-0012-2142]{T. G. Williams}
\affiliation{Max Planck Institute for Astronomy, K\"onigstuhl 17, 69117, Germany}

\author[0000-0003-0410-4504]{Ashley.~T. Barnes}
\affiliation{Argelander-Institut f\"{u}r Astronomie, Universit\"{a}t Bonn, Auf dem H\"{u}gel 71, 53121, Bonn, Germany}

\author[0000-0003-0166-9745]{F. Bigiel}
\affiliation{Argelander-Institut f\"ur Astronomie, Universit\"at Bonn, Auf dem H\"ugel 71, 53121 Bonn, Germany}

\author[0000-0002-5635-5180]{M\'elanie Chevance}
\affiliation{Universit\"{a}t Heidelberg, Zentrum f\"{u}r Astronomie, Institut f\"{u}r Theoretische Astrophysik, Albert-Ueberle-Stra{\ss}e 2, D-69120 Heidelberg, Germany}
\affiliation{Cosmic Origins Of Life (COOL) Research DAO, coolresearch.io}

\author[0000-0001-5301-1326]{Yixian Cao}
\affiliation{Max-Planck-Institut f\"ur Extraterrestrische Physik (MPE), Giessenbachstr. 1, D-85748 Garching, Germany}

\author[0000-0003-0085-4623]{Rupali Chandar}
\affiliation{Ritter Astrophysical Research Center, University of Toledo, Toledo, OH 43606, USA}

\author[0000-0002-5782-9093]{Daniel~A.~Dale}
\affiliation{Department of Physics and Astronomy, University of Wyoming, Laramie, WY 82071, USA}

\author[0000-0002-1185-2810]{Cosima Eibensteiner} 
\affiliation{Argelander-Institut für Astronomie, Universität Bonn, Auf dem Hügel 71, 53121 Bonn, Germany}

\author[0000-0001-6708-1317]{Simon C.~O. Glover}
\affiliation{Universit\"{a}t Heidelberg, Zentrum f\"{u}r Astronomie, Institut f\"{u}r Theoretische Astrophysik, Albert-Ueberle-Stra{\ss}e 2, D-69120 Heidelberg, Germany}

\author[0000-0002-3247-5321]{Kathryn Grasha}
\affiliation{Research School of Astronomy and Astrophysics, Australian National University, Canberra, ACT 2611, Australia}   
\affiliation{ARC Centre of Excellence for All Sky Astrophysics in 3 Dimensions (ASTRO 3D), Australia}   

\author{Stephen Hannon}
\affiliation{Department of Physics and Astronomy, University of California, Riverside, CA, 92521 USA}

\author[0000-0002-8806-6308]{Hamid Hassani}
\affiliation{Department of Physics, University of Alberta, Edmonton, Alberta, T6G 2E1, Canada}

\author[0000-0002-0432-6847]{Jaeyeon Kim}
\affiliation{Universit\"{a}t Heidelberg, Zentrum f\"{u}r Astronomie, Institut f\"{u}r Theoretische Astrophysik, Albert-Ueberle-Stra{\ss}e 2, D-69120 Heidelberg, Germany}

\author[0000-0002-0560-3172]{Ralf S.\ Klessen}
\affiliation{Universit\"{a}t Heidelberg, Zentrum f\"{u}r Astronomie, Institut f\"{u}r Theoretische Astrophysik, Albert-Ueberle-Stra{\ss}e 2, D-69120 Heidelberg, Germany}
\affiliation{Universit\"{a}t Heidelberg, Interdisziplin\"{a}res Zentrum f\"{u}r Wissenschaftliches Rechnen, Im Neuenheimer Feld 205, D-69120 Heidelberg, Germany}

\author[0000-0002-8804-0212]{J.~M.~Diederik~Kruijssen}
\affiliation{Cosmic Origins Of Life (COOL) Research DAO, coolresearch.io}

\author[0000-0001-7089-7325]{Eric J. Murphy}
\affiliation{National Radio Astronomy Observatory, 520 Edgemont Road, Charlottesville, VA 22903, USA}

\author[0000-0002-3289-8914]{Justus Neumann}
\affiliation{Max Planck Institute for Astronomy, K\"onigstuhl 17, 69117, Germany}

\author[0000-0002-1370-6964]{Hsi-An Pan}
\affiliation{Department of Physics, Tamkang University, No.151, Yingzhuan Road, Tamsui District, New Taipei City 251301, Taiwan} 

\author[0000-0003-3061-6546]{J\'{e}r\^{o}me Pety}
\affiliation{IRAM, 300 rue de la Piscine, 38400 Saint Martin d'H\`eres, France}
\affiliation{LERMA, Observatoire de Paris, PSL Research University, CNRS, Sorbonne Universit\'es, 75014 Paris}

\author[0000-0002-2501-9328]{Toshiki Saito}
\affiliation{National Astronomical Observatory of Japan, 2-21-1 Osawa, Mitaka, Tokyo, 181-8588, Japan}

\author[0000-0002-9333-387X]{Sophia K. Stuber}
\affiliation{Max Planck Institute for Astronomy, K\"onigstuhl 17, 69117, Germany}

\author[0000-0002-9483-7164]{Robin G. Tre{\ss}}
\affiliation{Institute of Physics, Laboratory for galaxy evolution and spectral modelling, EPFL, Observatoire de Sauverny, Chemin Pegais 51, 1290 Versoix, Switzerland.}

\author[0000-0003-1242-505X]{Antonio Usero}
\affiliation{Observatorio Astron\'{o}mico Nacional (IGN), C/Alfonso XII, 3, E-28014 Madrid, Spain}

\author[0000-0002-7365-5791]{Elizabeth~J. Watkins}
\affiliation{Astronomisches Rechen-Institut, Zentrum f\"{u}r Astronomie der Universit\"{a}t Heidelberg, M\"{o}nchhofstra\ss e 12-14, 69120 Heidelberg, Germany}

\author[0000-0002-3784-7032]{Bradley~C. Whitmore}
\affiliation{Space Telescope Science Institute, 3700 San Martin Drive, Baltimore, MD 21218, USA}

\collaboration{40}{(PHANGS)}

\begin{abstract}
Large-scale bars can fuel galaxy centers with molecular gas, often leading to the development of dense ring-like structures where intense star formation occurs, forming a very different environment compared to galactic disks. We pair $\sim$0.3\arcsec (30\,pc) resolution new JWST/MIRI imaging with archival ALMA CO(2-1) mapping of the central $\sim$5\,kpc of the nearby barred spiral galaxy NGC\,1365, to investigate the physical mechanisms responsible for this extreme star formation. 
The molecular gas morphology is resolved into two well-known bright bar lanes that surround a smooth dynamically cold gas disk ($\rm R_{gal} \sim 475\,pc$) reminiscent of non-star-forming disks in early type galaxies and likely fed by gas inflow triggered by stellar feedback in the lanes. The lanes host a large number of JWST-identified massive young star clusters. We find some evidence for temporal star formation evolution along the ring.
The complex kinematics in the gas lanes reveal strong streaming motions and may be consistent with convergence of gas streamlines expected there. Indeed, the extreme line-widths are found to be the result of inter-`cloud' motion between gas peaks; {\sc ScousePy} decomposition reveals multiple components with line widths of $\rm \langle\sigma_{CO,scouse}\rangle\,\approx\,19\,km\,s^{-1}$ and surface densities of $\rm \langle\,\Sigma_{H_2,scouse}\rangle\,\approx\, 800\,M_{\odot}\,pc^{-2}$, similar to the properties observed throughout the rest of the central molecular gas structure.  Tailored hydro-dynamical simulations exhibit many of the observed properties and imply that the observed structures are transient and highly time-variable.  From our study of NGC\,1365, we conclude that it is predominantly the high gas inflow triggered by the bar that is setting the star formation in its CMZ.
\end{abstract}


\section{Introduction} 
\label{sec:intro}

\begin{figure*}
    \centering
    \includegraphics[width=1.0\textwidth]{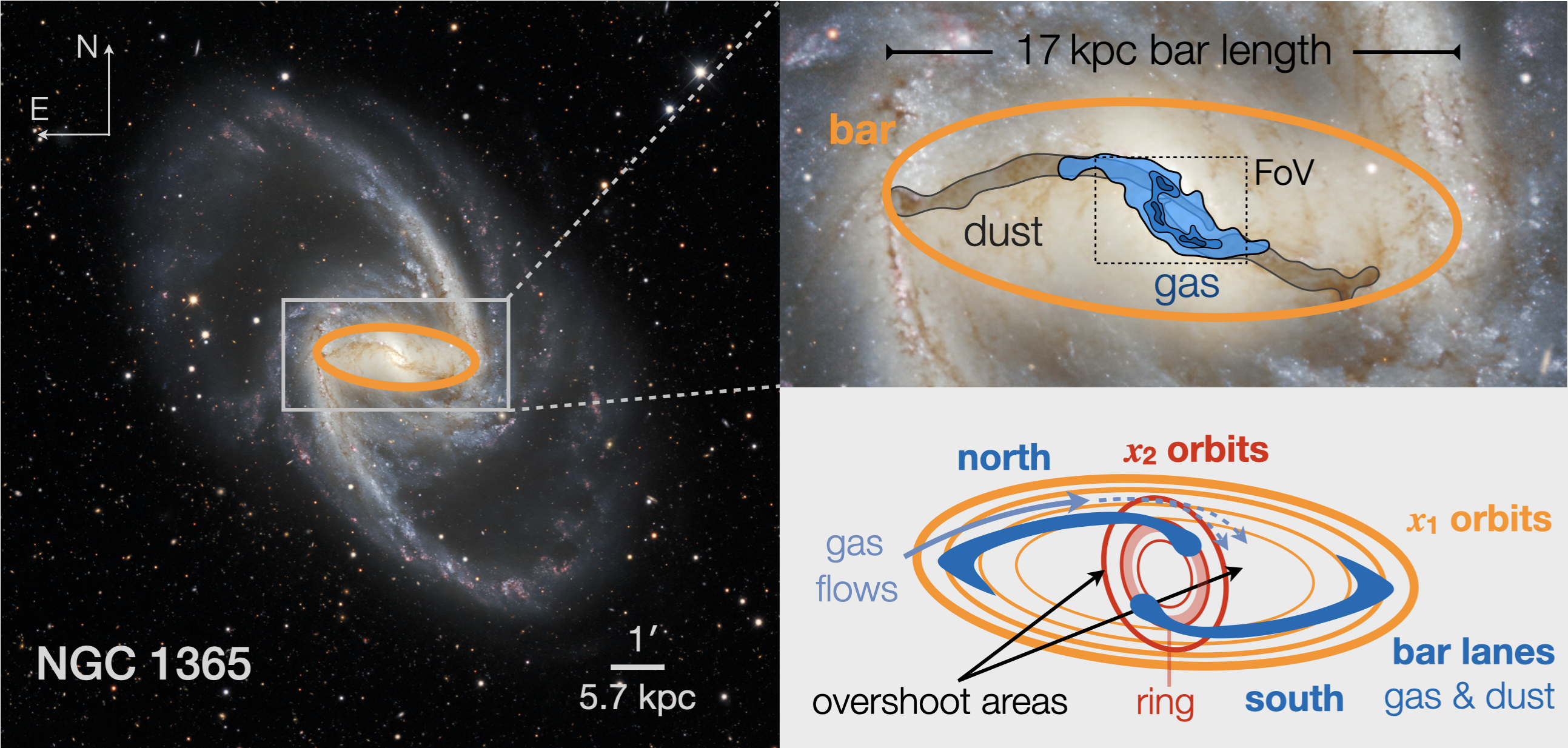}
    \caption{The inner 5\,kpc in the context of the large-scale stellar bar in NGC\,1365. The optical color image from DES\footnote{https://noirlab.edu/public/images/iotw2127a/} shows the 17\,kpc long stellar bar (outlined by the orange ellipse as determined by \citealt{herrera-endoqui2015}) surrounded by spiral arms (\textit{left}). In the zoom-in (marked as a grey box in the left panel), the bar lanes along the leading sides of the bar are highlighted via prominent extinction tracing dust (dark grey-shaded) and the molecular gas distribution (blue-shaded) (\textit{right top}). The field-of-view (FoV) under study in this work is shown as a dashed black box. The sketch (\textit{right bottom}) outlines the basic structure of orbits supporting the stellar bar ($x_1$ - elongated along the bar major axis, in orange; $x_2$ - elongated along the bar minor axis, in red) and the resulting bar lane morphology plus the inner star-forming ring. We further indicate the expected net gas flow, and some relevant regions discussed in this work.
    \label{fig:ngc1365_overview}
    }
\end{figure*}

Galaxy centers are special places for star formation to occur. They can contribute 10--100\% of the overall star formation in galaxies \citep{Kormendy2004}, capture extreme conditions and represent major sites of feedback to the circumgalactic medium \citep[e.g.,][]{Veilleux2020}. It is expected that star formation in galaxy centers proceeds differently than in disks. The dynamical time in galaxy centers is short (1-50\,Myr), with gas inflows driven by stellar bars (and spirals), large-scale gas outflows launched by active galactic nuclei (AGN) and central starbursts, and intense radiation fields due to the high stellar densities (and AGN when present). All of these phenomena affect the balance between the self-gravity of the molecular gas and factors (e.g., turbulence, shear, magnetic fields, tidal forces, cosmic ray flux) that support the gas against gravitational collapse \citep[e.g.,][]{chevance2020,Girichidis2020}. The closest galaxy centers allow for high physical resolution studies to assess the impact of these environmental factors making them unique targets for testing star formation theories \citep[for recent examples utilizing observations on molecular gas, see, e.g.,][]{Martin2021,Callanan2021,Levy2021,Eibensteiner2022,Behrens2022}.

The cold gas and dust distribution of the central regions of barred galaxies is shaped by the orbital structure of the underlying barred gravitational potential and typically forms two bar lanes along the leading sides of the bar which generally curve at smaller galactocentric radii \citep[Fig.\,\ref{fig:ngc1365_overview} right; see seminal paper by][]{Athanassoula1992}. In gas-rich galaxies, these inner regions (indicated as FoV in Fig.\,\ref{fig:ngc1365_overview}) often host a star-forming ring or spiral structure that are also referred to as nuclear rings/disks or more generally, in analogy to the center in the Milky Way, as central molecular zones (CMZs) \citep[e.g.][]{Morris1996,Sakamoto1999,Sheth2002,Sormani2015,Martin2021,henshaw2022}.

While these rings are often sites of intense star formation \citep{Knapen2000,comeron2014}, the physical processes leading to this massive star formation are far from understood with several models being proposed from gravitational instabilities in the ring \citep[e.g.][]{Elmegreen1994} or in dense spurs along the straight bar lanes \citep{Sheth2005} to gas collapse triggered at the location where gas from the lanes enters the ring \citep{boeker2008}. In the past years, several simulations have started to shed more light into the mechanisms leading to these rings. Some studies report a relation between gas mass inflow rate and resulting SFR in the ring \citep{Seo2013,Seo2019,Sormani2020,moon2021}. A varying inflow rate has been measured for the Galactic CMZ \citep{Sormani2019} and is also seen in simulations \citep[e.g.][]{Seo2019,tress2020}. Alternatively, quasi-periodic variations in the star formation activity are explained by the evolution of gas piling up in the ring (quiescent phase) and becoming gravitationally unstable (starburst phase) \citep[e.g.,][]{Loose1982,Krugel1993,Stark2004,Kruijssen2014,Emsellem2015,krumholz2017,Armillotta2019}.

With global simulations of gas flow in galactic centers now reaching pc or even sub-pc resolutions \citep[e.g.][]{renaud+2013,tress2020}, it is evident that comparisons to observations resolving similar spatial scales for the molecular gas and star formation are required to make progress in our understanding. Thanks to the advent of ALMA and JWST, it is now possible to probe the properties of molecular gas and (embedded) star formation at $\sim$0.2\arcsec\ which translates to 19\,pc at the distance of NGC\,1365 (19.6\,Mpc; \citealt{Anand2021a,Anand2021b}). In the PHANGS (Physics at High Angular resolution in Nearby GalaxieS\footnote{www.phangs.org}) sample, NGC\,1365 is the one barred galaxy that has already received JWST observations \citep[via PHANGS-JWST;][this issue]{LEE_PHANGSJWST} and has high-quality, high-resolution observations in the ALMA archive. Through the PHANGS set of surveys \citep[][]{Leroy2021-Survey,Emsellem2022,Lee2022} abundant complementary information on this galaxy is available \citep[e.g.][]{sun2022}.

NGC\,1365 is a nearby ($\rm D=19.6\,Mpc; 1\arcsec\approx 95\,pc$; \citealt{Anand2021a,Anand2021b}) barred spiral galaxy hosting an AGN \citep[][]{Morganti1999} and has the highest SFR in the PHANGS-ALMA sample \citep{Leroy2021-Survey}.
\cite{Lindblad1999} provides a detailed review on NGC\,1365 and we summarize its basic parameters in Tab.\,\ref{tab:ngc1365}. Located in the Fornax cluster \citep[][]{Jones1980}, this grand-design spiral galaxy is morphologically classified as (R')SB(r{\underline s},nr)bc, indicating the presence of an outer pseudoring, a weak inner pseudoring corresponding to the surroundings of the bar, and a prominent circumnuclear ring \citep[][]{Buta2015}. The bar is remarkably long, with a full length of 17.2\,kpc in the plane of the sky \citep[][]{herrera-endoqui2015}, which corresponds to a deprojected value of ${\sim}28$\,kpc in the plane of the galaxy\footnote{This deprojection relies on the kinematic parameters from Table\,\ref{tab:ngc1365} and follows a simple deprojection considering a ``1D'' bar \citep[e.g.][]{Martin1995}.}.
The nucleus harbors a number of compact radio sources and a large number of super star clusters, with star formation taking place mostly in an elongated circumnuclear ring \citep[e.g.,][]{Kristen1997,Forbes1998,Stevens1999,Galliano2005,Alonso-Herrero2012,Fazeli2019}, which likely corresponds to the inner Lindblad resonance at $r \approx 1$\,kpc reported by \cite{Lindblad1996}. This ring is very rich in molecular gas \citep[][]{Sandqvist1995,Sakamoto2007,Gao2021,Egusa2022}.
The AGN is known to drive a biconical outflow seen in ionized gas \citep[e.g.,][]{Storchi-Bergmann1991,Veilleux2003,Venturi2018}. All these properties make NGC\,1365 an ideal target for a detailed study of the CMZ properties utilizing ALMA and JWST observations.

The letter is organized as follows. After a brief description of the data and simulations used (\S\,\ref{sec:data}), we present the inferred properties of the gas and star formation in the starburst ring (\S\,\ref{sec:results}) and discuss them also in context of the simulations in \S\,\ref{sec:discussion}. We present our conclusions towards improving our understanding of CMZs in \S\,\ref{sec:conclusion} before summarizing our findings (\S\,\ref{sec:summary}). 

\begin{deluxetable}{lcc}
\tablecaption{Parameters for NGC\,1365\label{tab:ngc1365}}
\tablehead{
\colhead{Parameter} &
\colhead{Value} &
\colhead{Reference}}
\startdata
R.A. (J2000) & 03h33m36.37s & (1) \\
Dec. (J2000) & -36d08m25.4s & (1) \\
$\rm v_{sys}$ (LSR) & $\rm 1613\pm5\,km\,s^{-1}$ & (2) \\
D & $\rm 19.6\pm0.8\,Mpc$ & (3), (4) \\
Incl. & $\rm 55.4\pm6.0^{\circ}$ &  (2) \\
P.A. & $\rm 201.1\pm7.5^{\circ}$ & (2) \\
$\rm M_{\star}$ & $\rm log(M_{\star}/M_{\odot})=11.0\pm0.2$ & (1) \\
SFR & $\rm log(SFR/M_{\odot}yr^{-1})=1.24\pm0.2$ & (1)\\
$\rm M_{H_2}$ & $\rm log(M_{H_2}/M_{\odot})=10.3$ & (1)\tablenotemark{a}\\
$\rm SFR_{starburst}$ & $\rm log(SFR/M_{\odot}yr^{-1})\approx 0.7$ & (5)\tablenotemark{b}\\
$\rm M_{H_2,starburst}$ & $\rm log(M_{H_2}/M_{\odot})\approx 10.0$ & (1)\tablenotemark{c}\\
Bar PA & $86^{\circ}$ & (6)\\
Bar radius\tablenotemark{d} & 90.4\arcsec\ (8.6\,kpc) & (6)\\
\enddata
\tablenotetext{a}{$\rm H_2$ gas mass assuming a standard conversion factor and $R_{21}=1$ (see \S\,\ref{subsec:alma}) including aperture correction.}
\tablenotetext{b}{SFR of the starburst (i.e. inside $\rm R_{gal} \le 1.8\,kpc$) as measured on the attenuation corrected H$\alpha$ map from PHANGS-MUSE \citep{Emsellem2022}.}
\tablenotetext{c}{$\rm H_2$ gas mass of the starburst (i.e. inside $\rm R_{gal} \le 1.8\,kpc$).}
\tablenotetext{d}{projected radius \citep{herrera-endoqui2015}; deprojected bar radius is $\sim$14\,kpc using the orientation provided in the table.}
\tablecomments{The parameters are taken from:
(1) \cite{Leroy2021-Survey}, 
(2) \cite{lang2020},
(3) \cite{Anand2021a},
(4) \cite{Anand2021b},
(5) \cite{Belfiore2022}, 
(6) \cite{herrera-endoqui2015}
}
\end{deluxetable}

\section{Data} 
\label{sec:data}

With a SFR of $\rm 5\,M_{\odot}yr^{-1}$, the massive starburst in the center of the strongly barred galaxy NGC\,1365 makes it a prime candidate for the study of how star formation proceeds in `circum-nuclear rings' or CMZs. For this, we combine archival ALMA imaging (\S\,\ref{subsec:alma}) with new JWST/MIRI imaging sensitive to the embedded phase of star formation (\S\,\ref{subsec:jwst}) at matched resolution of $\sim$0.3\arcsec\ and compare to spectroscopic H$\alpha$ imaging at lower resolution (\S\,\ref{subsec:muse}). Here we also briefly describe the set-up of the simulations (\S\,\ref{subsec:ramses}) used for comparison.

\subsection{ALMA data}
\label{subsec:alma}

The molecular gas distribution in the central 225\arcsec$\times$125\arcsec\ (PA~$\sim-5^{\circ}$) of NGC\,1365 has been observed in its CO(2-1) line using the ALMA 12m together with the ACA (7m array and total power antennas) as part of project 2013.1.01161.S (PI: K. Sakamoto) in Cycle 2. \cite{Leroy2021-Survey} presented the calibration and reduction for the combined data from the two more compact 12m array configurations (C34-1 \& C34-3) and the ACA. The observations with the extended 12m configuration (C34-5) only cover the central 80\arcsec$\times$50\arcsec\ (PA~$\sim-30^{\circ}$) which corresponds to roughly 7.6$\times$4.8\,kpc$^2$. We applied the observatory delivered calibration (\textit{scriptForPI.py}). The PHANGS-ALMA imaging pipeline \citep{Leroy2021-Pipeline} was adapted and used to simultaneously image the three 12m array configurations together with the 7m array data. In a next step the total power data was feathered with the interferometric data cube as outlined in \cite{Leroy2021-Pipeline}. The resulting cube has a spectral resolution of about 2.54\,km\,s$^{-1}$, an angular resolution of $\sim$0.31\arcsec\ which corresponds to $\sim$30\,pc at our adopted distance of $19.57 \pm 0.78$\,Mpc \citep{Anand2021a,Anand2021b}, and an rms of 0.72\,K.

The resulting data products are derived with the PHANGS-ALMA imaging pipeline as described by \cite{Leroy2021-Pipeline}. For the analysis, we utilize the following products: integrated intensity map using the broad mask (i.e.\ high completeness), velocity field derived with prior, peak temperature map obtained in $\rm 12.5\,km\,s^{-1}$ wide channels, and map of effective velocity width. Differences between our integrated CO(2-1) intensity map and that presented by \cite{alonso-herrero2020} are due to the fact that the latter was derived from the extended 12m array configuration data only.

In order to convert CO(2-1) intensities $\rm I_{CO21}$ to H$_2$ gas mass surface densities $\rm \Sigma_{H_2}$, we adopt a Galactic conversion factor of $\rm \alpha_{CO}=4.35\,M_{\odot}\,pc^{-2}\,(K\,km\,s^{-1})^{-1}$ \citep{bolatto2013}, and a CO(2-1) to CO(1-0) ratio of $R_{21}=1$. \citet[][this issue]{LEROY2_PHANGSJWST} recommend this ratio for high MIR surface brightness (e.g. $\rm I_{21\mu m} \geq 10\,MJy\,sr^{-1}$ as present here) based on an extensive comparison of CO and MIR emission in nearby galaxies. This is close to the $R_{21}$ values around 0.8-0.9 obtained at $\sim$3\arcsec\ resolution for the region studied here \citep[][this issue]{LIU_PHANGSJWST}. \cite{Teng2022} report a lower $\rm \alpha_{CO}\approx0.5-2.0\,M_{\odot}\,pc^{-2}\,(K\,km\,s^{-1})^{-1}$ for the starburst ring in the nearby barred galaxy NGC\,3351 based on non-LTE modeling of multiple CO lines. Hence H$_2$ surface densities could be a factor 2-4$\times$ lower than quoted here depending on the exact conditions of the molecular ISM \citep[see also detailed analysis for NGC\,1365's central region by][]{LIU_PHANGSJWST}.

\subsection{JWST/MIRI and NIRCam Imaging}
\label{subsec:jwst}

As part of the PHANGS-JWST Treasury program \citep[project ID 2107,][this issue]{LEE_PHANGSJWST}, JWST has mapped NGC~1365 using the MIRI instrument in a $2\times2$ mosaic in four filters -- F770W, F1000W, F1130W, and F2100W, and a $2\times1$ mosaic using NIRCam in an additional four filters (F200W, F300M, F335M, and F360M), the most prescient for this work being the F200W image. Each of these mosaic tiles use a four-point dither pattern, to ensure good sampling of the PSF (0.066/0.243/0.321/0.368/0.665\arcsec at F200W/F770W/F1000W/F1130W/F2100W). Full details of the data processing are given in \cite{LEE_PHANGSJWST}, but as a brief overview we use the public JWST pipeline\footnote{\url{https://github.com/spacetelescope/jwst}} mostly with default settings, with the latest reference files at the time of processing (early September 2022, although we use the improved MIRI flats delivered mid-September; K. Gordon, priv. comm.). These latest files improve upon the mosaics that were available from the MAST archive. 
For the MIRI imaging, we use dedicated background observations of the galaxy to remove any thermal background from the observations, as recommended by the observatory \citep{Rigby2022}. The thermal background is negligible in the NIRCam imaging, so this step is not performed for the F200W data. We found that the simultaneously recorded data of MIRI's Lyot coronagraph has a noticeably different background to the main science detector, and mask this out before mosaicking. Given the outstanding background uncertainties, we currently image each of the MIRI fields separately and mosaic them together outside of the JWST pipeline, but we expect the background matching may improve as the pipeline is updated. Absolute astrometric alignment is based on cross-correlation of the MIRI mosaics with an already-aligned NIRCam image taken as part of the same observations, which uses AGB stars in the galaxy detected in HST imaging \citep[for details, see][]{LEE_PHANGSJWST}. The final NIRCam F200W mosaic has a sensitivity of around $\rm 0.1\,MJy\,sr^{-1}$, and the final MIRI mosaics have sensitivities that vary from around $\rm 0.1\,MJy\,sr^{-1}$ for F770W to $\rm 0.3\,MJy\,sr^{-1}$ for F2100W. Finally, as our images are filled with galaxy emission, to achieve an absolute flux level we anchor the fluxes to existing {\it Spitzer} or WISE imaging, deriving a constant offset for each band \citep[see Appendix A and B of][this issue]{LEROY2_PHANGSJWST}. We perform no convolution to the images, to maximize the resolution of each MIRI image.


\subsection{PHANGS-MUSE Data Products}
\label{subsec:muse}

We make use of data from the PHANGS-MUSE survey \citep[PI Schinnerer; ][]{Emsellem2022} which imaged significant parts of the star-forming disk using the optical integral field unit (IFU) MUSE mounted on the VLT. Data processing and generation of ionized gas emission line maps follow standard procedures and are described in detail in \cite{Emsellem2022}. We utilize the H$\alpha$ emission line map from the \textit{copt} (i.e. convolved to a common resolution across the spectral range and all pointings) 1.15\arcsec\ resolution mosaic with a sensitivity of $\rm 2\times\,10^{-20}\,erg\,s^{-1}\,cm^{-2}\,spaxel^{-1}$ for 0.2\arcsec\ spaxels. In addition, the attenuation-corrected H$\alpha$ map (assuming $R_V = 3.1$) and corresponding E(B-V) map (from the Balmer decrement; assuming case B recombination, temperature $T=10^4\,K$, density $n_e=10^2\,\mathrm{cm}^{-3}$) derived by \cite{Belfiore2022} from the PHANGS-MUSE \textit{copt} data products are used. 

\subsection{{\sc Ramses} Hydro-dynamical Simulations of NGC\,1365}
\label{subsec:ramses}

In the course of the PHANGS project, we designed and ran dedicated hydrodynamical simulations tuned to mock the overall properties of a sub-sample of PHANGS targets: we briefly present here some results from the NGC\,1365-like simulation (see Emsellem et al., in preparation, for a more extensive description). To set up the simulations, we made use of NGC\,1365's known observed global properties (see Tab.\,\ref{tab:ngc1365}) and radial profiles, namely the molecular and H{\sc i} content, stellar mass, stellar density profile and CO velocity profile \citep{sun2022}, together with geometrical considerations (inclination, PA of the line of nodes), to construct a multi-component axisymmetric three-dimensional mass and dynamical model including stars, dark matter and gas, using the Multi-Gaussian Expansion formalism \citep[MGE;][]{emsellem+1994,emsellem+2015} as in, e.g., \cite{renaud+2013}. We then conducted a non-cosmological adaptive mesh refinement hydrodynamical simulation using the {\sc Ramses} code \citep{teyssier2002}, starting with initial conditions based on a realization of that model, with live dark matter and stellar particles over a grid of 120\,$\times$120\,kpc$^2$, with a maximum sampling for the gas cells of $\sim 3.7$~pc (maximum refinement level of 15). We adopt sub-grid prescriptions for the cooling, star formation, feedback and stellar evolution, as in \cite{renaud+2021}. In short, the simulation accounts for atomic and molecular cooling, heating from an external UV flux, and star formation at a constant efficiency per free-fall time (2\%)  in dense gas (150~cm$^{-3}$), with initial masses of new particles of 2000~\Msun. It also includes prescriptions for stellar winds, radiative pressure, type-II and Ia supernova feedback with energy released based on the resolution of the local cooling radius \citep{agertz+2015, agertz+2021}. The NGC\,1365-like simulation was run for about 6.5~Gyr keeping the gas warm ($5\times10^4$~K) and isothermal to let the bar structure develop itself (and save CPU time), then followed by more than 1~Gyr of evolution with all subgrid recipes turned on (i.e., cooling, star formation) at maximum resolution.

In the simulation, a first-generation 8~kpc bar develops over the first Gyr, which then weakens between 1 and 2 Gyr, slowly regrowing at later times to reach about 20\,kpc length after 6\,Gyr. The resulting large-scale bar exhibits an extended central mass concentration of old and young stars, together with large-scale gas lanes wrapping around the central few kpc, reminiscent of what is observed for NGC\,1365. For comparison to the observations we select by visual inspection a snapshot that broadly resembles the gas intensity and line width distribution in NGC\,1365's inner 5\,kpc (see \S\,\ref{subsec:simulations}). We derived the fractional H$_2$ mass per cell from the simulation in post-processing by applying the gas density and metallicity dependent prescription from \citet{krumholz+2009}.

\section{Results}
\label{sec:results}

As we aim to study the circumnuclear star formation process, we focus our analysis on the inner $\sim$5\,kpc region of NGC\,1365. The analysis of the molecular gas emission encompasses the standard moment maps (\S\,\ref{subsec:COring}) and a kinematic analysis including a parametric decomposition of the gas emission using {\sc ScousePy} (\S\,\ref{subsec:scouse}). The study of the dust emission relies mostly on the JWST/MIRI imaging in the F770W and F1000W filters probing emission from PAHs and (very) small grains which trace the cold ISM and young (embedded) star-forming regions (\S\,\ref{subsec:mir-data}).

\subsection{Properties of the Molecular Gas in NGC\,1365's Inner Disk}
\label{subsec:COring}

\begin{figure*}
    \centering
    \includegraphics[width=1.0\textwidth]{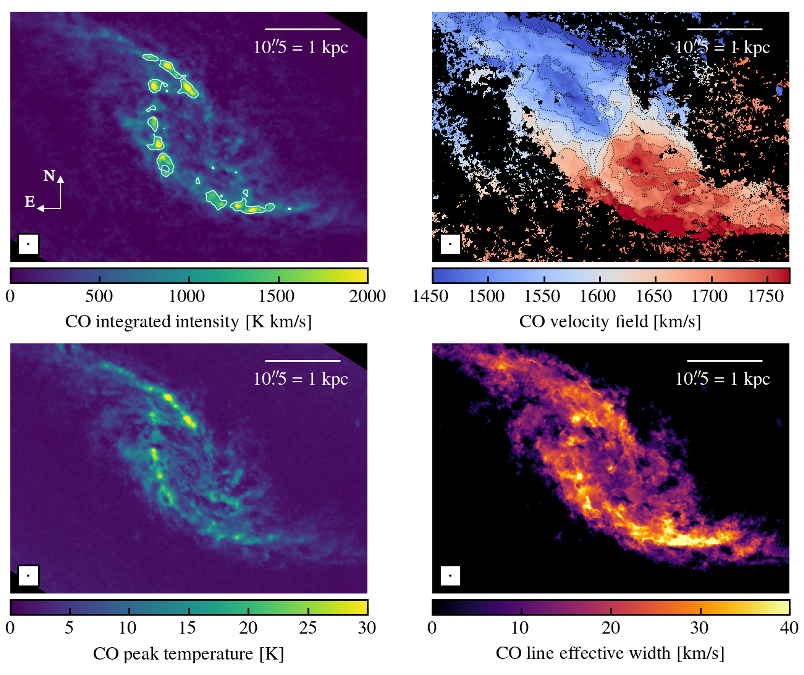}
    \caption{Molecular gas as probed by the CO(2-1) line emission from the inner $\rm 4.8\times3.3\,kpc^2$ of NGC\,1365 at $\sim$30\,pc (0.3\arcsec) resolution observed by ALMA. \textit{Top:} The integrated intensity distribution (\textit{left}) reveals a faint smooth gas disk inside the prominent gas spiral arms. The white contour highlights $\rm I_{CO}=1000\,K\,km\,s^{-1}$. The velocity field (\textit{right}) shows circular rotation in the inner disk and clear deviations from this pattern at the locations of the gaseous bar lanes. The iso-velocity contours are in steps of $\rm 40\,km\,s^{-1}$ from the systemic velocity.
    \textit{Bottom:} The CO(2-1) peak temperature map (\textit{left}) reveals several peaks reaching temperatures above 20\,K which often agree with peaks in the integrated intensity. Enhanced effective widths (\textit{right}) roughly coincide with the brightest emission in the bar lanes. The faint emission peak associated with the nucleus shows larger line widths above 40\,km\,s$^{-1}$.
    \label{fig:ngc1365_co}
    }
\end{figure*}

We show the integrated intensity map (broad mask), the peak temperature map ($T_\mathrm{peak}$), the velocity field ($v_\mathrm{LSR}$), and effective width (EW) distribution of the CO(2-1) line emission at $\sim$30\,pc (0.3\arcsec) resolution in Fig.\,\ref{fig:ngc1365_co}. The effective width is the inferred velocity dispersion from the integrated intensity and peak temperature if the line profile were Gaussian: $EW = I_\mathrm{CO} / (\sqrt{2\pi} T_\mathrm{peak})$.
CO(2-1) line emission is detected well above 5$\sigma$ from the inner roughly 50\arcsec$\times$35\arcsec\ ($\rm 4.8\times 3.3\,kpc^2$). The gas emission is resolved into a bright spiral-like structure that corresponds to the inner ends of the bar lanes seen in molecular gas and dust along the large-scale (28\,kpc in diameter) stellar bar (see Fig.\,\ref{fig:ngc1365_overview}). These bar lanes encompass a lower brightness disk inside a galactocentric radius of $\rm R_{gal}\approx$5\arcsec\ that only becomes evident at this resolution. At the location of the nucleus there is a weak compact peak with an integrated intensity of $\rm I_{CO}\approx790\,K\,km\,s^{-1}$. 
The southern bar lane reaches the northern one while the northern lane fades at a galactocentric radius of R$\sim$10\arcsec\ before reaching the southern one. Although no pronounced lane structure
is visible northwest of the nucleus, fainter CO emission is present.
While the overall gas morphology is similar to that seen at a lower resolution of $\sim$2\arcsec\ in various low-J CO transitions \citep[e.g.,][]{Sakamoto2007,Gao2021}, the bright emission breaks up into about 13 and 5 prominent peaks above an integrated surface brightness of $\rm I_{CO}\ge1000\,K\,km\,s^{-1}$ and sizes of $\sim$0.5-1\arcsec\ in the southern and northern gas lane, respectively. The peak brightness temperature in these integrated emission peaks is mostly well above $T_{\rm peak}=15$\,K. Interestingly, a set of very compact peaks in $T_{\rm peak}$ becomes apparent southwest of the nucleus (but north of the southern bar lane).

The velocity field of the inner low-surface brightness disk is very reminiscent of a rotating disk with a position angle close to the value of $\sim200^{\circ}$ inferred from the large-scale CO(2-1) velocity field \citep{lang2020}. Where the bar lanes reach smaller galactocentric radii, the velocity field significantly deviates from circular rotation, implying strong streaming motions (i.e. non-circular motions in the plane of the galaxy) (Fig.\,\ref{fig:ngc1365_co}). Interestingly, there is no clear or simple correlation between molecular gas surface brightness and velocity field deviation along the bar lanes. This behaviour was already visible in the 2\arcsec\ lower resolution data \citep{Gao2021}. However, the abrupt change of the iso-velocity contours at the location of---in particular---the southern lane becomes much more evident at 0.3\arcsec\ resolution.

The $\sim$30\,pc resolution data reveal a low mean effective line width of $\rm \langle EW \rangle=14\,km\,s^{-1}$ inside $\rm R_{gal}\approx$5\arcsec, i.e.\ for the inner circular rotating gas disk, with a $\sim$3$\times$ higher value at the very nucleus. At the location of the prominent gas lanes EW increases significantly and reaches values well above $\rm 30\,km\,s^{-1}$ particularly in the southern bar lane. High EW regions are often coincident with the compact integrated brightness peaks, but not always. The mismatch is particularly apparent for the northern bar lane.

\subsection{Distribution of PAH and Hot Dust Emission}
\label{subsec:mir-data}

\begin{figure*}
    \centering
    \includegraphics{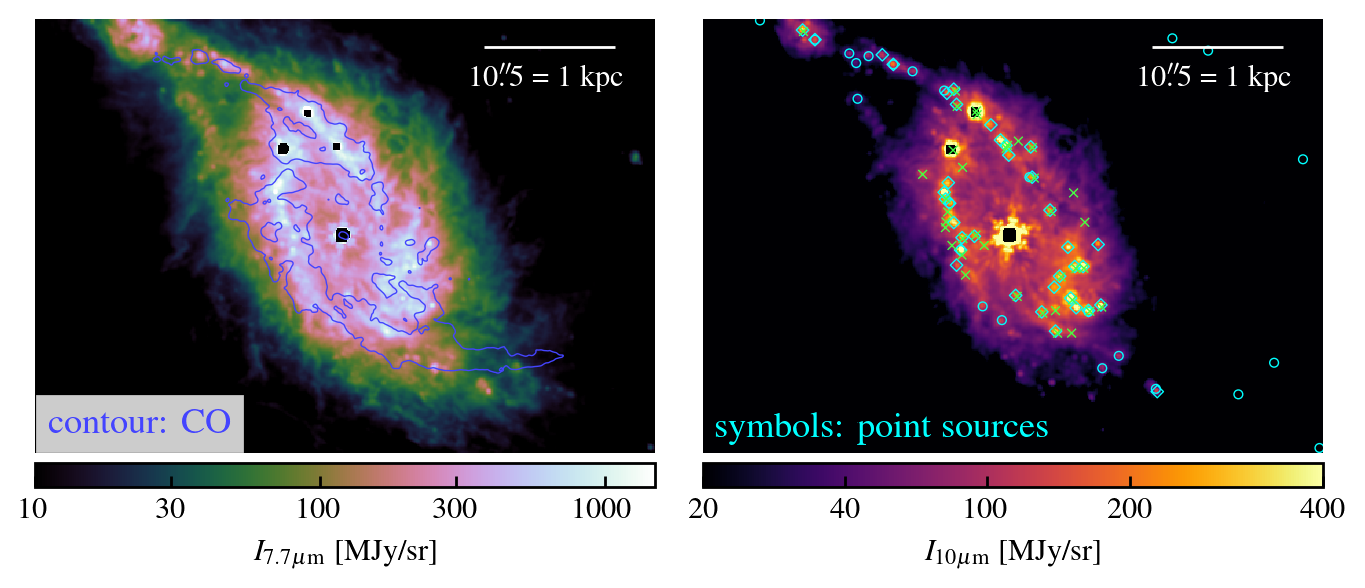}
    \caption{Distribution of the emission from PAHs and hot dust emission as probed by JWST/MIRI observations at $\sim$0.3\arcsec\ resolution, which is comparable to the resolution of the ALMA CO(2-1) data (North is up, East to the left.). 
    \textit{Left:} The F770W filter with a $\sim$0.24\arcsec\ resolution reveals the overall distribution of the neutral ISM as its emission is dominated by the  7.7$\mu$m  PAH feature. For reference the distribution of the CO emission is shown by a single blue contour. Note that the central position is significantly affected by saturation due to the bright AGN and dominated by its PSF causing the artefacts. Also the three brightest compact sources north of the AGN are saturated.
    \textit{Right:} Several embedded young star-forming clusters are evident in the 10\,$\mu$m emission and stand out as bright compact sources. Our 10\,$\mu$m-selected sources are highlighted by the cyan diamond symbols and are only found in the region shown. About 57\% of these objects coincide with the young ($<$10\,Myr) star clusters studied by \citet[][shown as green crosses]{WHITMORE_PHANGSJWST} and have estimated stellar masses of $\rm 10^6\, M_{\odot}$ or higher. The 21\,$\mu$m-selected sources from \cite{HASSANI_PHANGSJWST} that are classified as ISM emitting sources are shown as cyan circles. The dearth of clusters in the bar lane southwest of the nucleus is notable.}
    \label{fig:jwst} 
\end{figure*}

The JWST/MIRI broad-band filters are excellent tracers of the distribution of the ISM as they probe emission from very small dust grains (heated by single photons), PAHs, hot thermal dust continuum and, to a lesser degree, light from stars luminous in the near-IR, such as AGB stars \citep[e.g., review by][]{galliano2018}. 

In particular, the F770W filter encompasses emission from the 7.7$\mu$m PAH feature and provides a detailed view of the distribution and morphology of the neutral ISM at its $\sim$0.24\arcsec\ ($\sim$23\,pc) resolution \citep[e.g.,][this issue]{SANDSTROM1_PHANGSJWST,LEROY1_PHANGSJWST}. 
The F770W distribution in NGC\,1365 reveals a prominent central disk with filamentary morphology that extends beyond the molecular bar lanes
(Fig.\,\ref{fig:jwst}, left). It is interesting to note that only a few shell-like features or bubbles (with diameters of 10 to a few 100\,pc) are obvious, e.g., west of the northern bar lane. This is in stark contrast to the central 3\,kpc of the nearby grand-design spiral galaxy NGC\,628 where \citet[][this issue]{WATKINS_PHANGSJWST} visually identified 569 (presumably stellar feedback-driven) bubbles with a mean diameter of 77\,pc. This might suggest that the PAH distribution in NGC\,1365's center is not shaped by stellar feedback. The brightest emission ($\rm I_{F770W} > 400\,MJy\,sr^{-1}$) forms a ring-like distribution between galactocentric radii of $\rm R\approx 5-13\arcsec$ that roughly coincides with the CO(2-1) emission. However, there are notable differences between the PAH and CO distributions: 
(a) southwest of the nucleus, bright PAH emission `fills' the ring while bright CO emission is absent,  
(b) the southern CO bar lane outside a galactocentric radius of $\rm R\sim\,13\arcsec$ exhibits bright to very bright CO emission without correspondingly bright PAH emission, and
(c) a linear fainter PAH emission feature about 15\arcsec\ to the north-east of the nucleus connects the ring to the northern bar lane without a counterpart in CO emission. 

\smallskip
Among the filters used by PHANGS-JWST, F2100W is the most sensitive to emission from hot dust as it probes the longest wavelength range where emission from PAHs no longer dominate \citep[also evident from its different scaling to CO emission, see][]{LEROY1_PHANGSJWST}. Due to its lower resolution of $\sim$0.67\arcsec\ and significant saturation affecting several regions in the central 20\arcsec\ of NGC\,1365, we resort to the F1000W map which is less affected by saturation. This filter probes the underlying dust continuum in star-forming galaxies but can be affected by silicate absorption which is mostly seen in AGN and MIR-spectra of ULIRGs \citep{galliano2018,spoon2007}. It typically probes dust heated by single photons \citep[see review by][]{galliano2018}, although the analysis by  
\citet[][]{LEROY1_PHANGSJWST} suggests a significant contribution of PAH emission especially at lower intensities. 
The overall distribution of the 10\,$\mu$m emission (Fig.\,\ref{fig:jwst}, right) is as expected similar to that of the 7.7\,$\mu$m PAH emission. 

In addition, several compact bright sources are evident. We generate a catalog of bright compact sources identified in the F1000W image following the methodology outlined in \citet[][this issue]{HASSANI_PHANGSJWST}. They select bright compact sources at 21$\mu$m, show that 85\% are consistent with being embedded star-forming regions (the remainder are background galaxies ($\sim$10\%) or dusty stars ($\sim$5\%)), and suggest that the 10$\mu$m emission of these objects is consistent with probing hot dust emission.
Selecting only 10$\mu$m-identified sources with $\rm F_{F1000W} > 100\,\mu Jy$ and $\rm F_{F2100W} > 200\,\mu Jy$, we obtain a total of 37 compact objects (see diamond symbols in Fig.\,\ref{fig:jwst}, right). Comparison to ground-based N-band imaging work by \citet{Galliano2005} reveals that all of their seven off-nuclear mid-IR sources are detected in the JWST map with sources M5 and M6 being saturated there (and added by hand to our 10\,$\mu$m sample).

To assess the completeness of dusty star-forming regions, we compare our 10\,$\mu$m-selected sample to the 21\,$\mu$m-selected embedded cluster candidates from \cite{HASSANI_PHANGSJWST}\footnote{Their sources with an `\texttt{ISM\_EYE}=true' flag, i.e. they are being visually classified as having their 21\,$\mu$m emission due to ISM/hot dust and not being a background galaxy or a dusty star.}. There are eighteen embedded star-forming cluster candidates with $\rm \nu L_{\nu,21\mu m} = 10^6-10^8\,L_{\odot}$ present in the molecule- and PAH-bright region\footnote{Another six sources can be seen well outside this region in the West and are disregarded in the further discussion.}. Eight candidates ($\sim$45\%) coincide with our 10\,$\mu$m-selected sources. All remaining ten sources are identified at 10\,$\mu$m by our method albeit at flux levels below our imposed flux cut, implying that their mid-IR colors (21\,$\mu$m-10\,$\mu$m) are redder. The brightest mid-IR sources like M4, M5, M6 and likely a few others southwest of the nucleus are missing from the 21\,$\mu$m selection due to saturation. 

\cite[][this issue]{WHITMORE_PHANGSJWST} have compiled a sample of 37 young ($\rm t_{age} \leq 10\,Myr$), massive ($\rm M_{\star} \geq 10^6\,M_{\odot}$)
star clusters based on a combination of optical \citep[from HST; ][Whitmore et al. subm.]{turner2021}, radio \citep[][]{Sandqvist1995}, ground-based \citep{Galliano2005} and space-based (JWST filters: F335M, F770W, F1130W) mid-IR observations. The mean age, stellar mass, and attenuation of these sources are
$\rm log(t_{age}/yr)=6.5\pm0.2$, $\rm log(M_{\star}/M_{\odot})=6.3\pm0.3$ and $\rm A_V=6.3\pm3.4\,mag$ based on estimates from a combination of HST and JWST information \citep[for details see][]{WHITMORE_PHANGSJWST}.
About twenty-one of these clusters (57\%) are in common with our 10\,$\mu$m-selected sources (when including the saturated M5 and M6 sources). Strongly supporting our notion that all 10\,$\mu$m-identified sources are also likely massive young clusters (Fig.\,\ref{fig:jwst} right). The three brightest clusters (all badly saturated in the F2100W filter) are M4, M5, and M6 which are also the three strongest radio continuum sources at 6\,cm \citep{Sandqvist1995}. 
For the brightest off-nuclear sources at 10\,$\mu$m M4, M5, and M6, \citet{galliano2008} quote ages of $\sim$7\,Myr and stellar masses of $\rm \sim10^7\,M_{\odot}$ based on MIR spectroscopy, while for the same objects \citet{galliano2012} report ages of $\rm log(t_{age}/yr)\sim 5.5-6.5$ and stellar masses of $\rm log(M_{\star}/M_{\odot})\sim 6.5-7.0$ based on NIR integral field spectroscopy which are consistent with the values of \citet[][]{WHITMORE_PHANGSJWST} of $\rm log(t_{age}/yr)=6.5/6.5/6.5$ and $\rm log(M_{\star}/M_{\odot})=6.0/6.6/6.7$ for M4, M5, M6, respectively. These discrepancies demonstrate the challenge of obtaining accurate ages for these clusters.

The union of these three catalogs should provide a good representation of the distribution of ongoing massive star formation in the CMZ of NGC\,1365 independent of the amount of dust being present. The distribution of these young (embedded) cluster candidates falls broadly into three categories:
(i) delineating very well the gaseous bar lanes northeast of the nucleus (including the MIR-brightest M4, M5, M6 clusters) with a fair number of JWST-only identified clusters along the northern lane,
(ii) a more stochastic distribution south-west of the nucleus, and
(iii) a remarkable dearth of clusters in the outer southern CO lane.

\subsection{Molecular Gas Dynamics in Inner 5\,kpc}
\label{subsec:scouse}

\begin{figure*}
    \centering
    \includegraphics[trim=0cm 15cm 0cm 2cm, clip, width=1.0\textwidth]{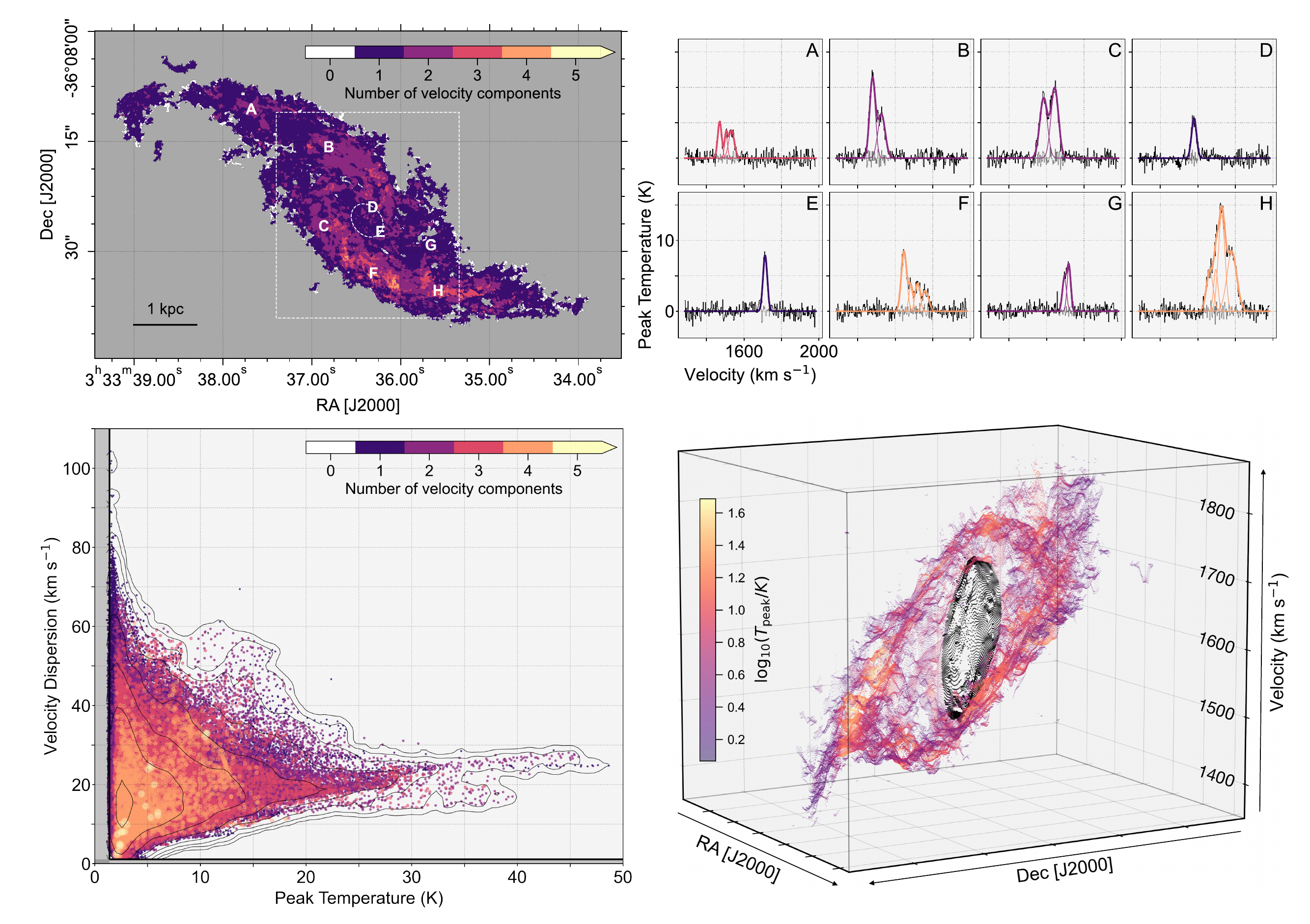}
    \caption{{\sc ScousePy} decomposition of the 30\,pc resolution molecular CO(2-1) gas emission. Multiple Gaussian components are needed to fit $\sim31\%$ of the data. This is demonstrated most clearly in the spiral arms, where two and three component models are common (up to six components in the southwestern arm; \textit{top left}). Spectra from selected regions with the resulting {\sc ScousePy} fits typically exhibit well separated line components (\textit{top right}). The inferred peak CO temperature and velocity dispersions of the individual {\sc ScousePy} components (\textit{bottom left}) occupy a similar region of parameter space independent of whether they are derived from single or multiple components fits (color-coding). The grey shaded regions here correspond to the limits of our fitting (a signal-to-noise ratio of 2 and a single channel width).
    The three-dimensional representation of the individual {\sc ScousePy} components located within the white dashed box in the top left panel, displayed in $ppv$ space (\textit{bottom right}), reveals an inner smoothly rotating disk (highlighted in black; extracted from the ellipse in the top left panel) while the bar lanes show strong local fluctuations in velocity indicating an increasing complexity of the kinematics in the molecular gas there (the color here refers to the peak CO temperature of the individual components). The physical spacing of the $pp$ grid in this image is $\sim475$\,pc.}
    \label{fig:scouse} 
\end{figure*}

\begin{figure*}
    \centering
    \gridline{
    \fig{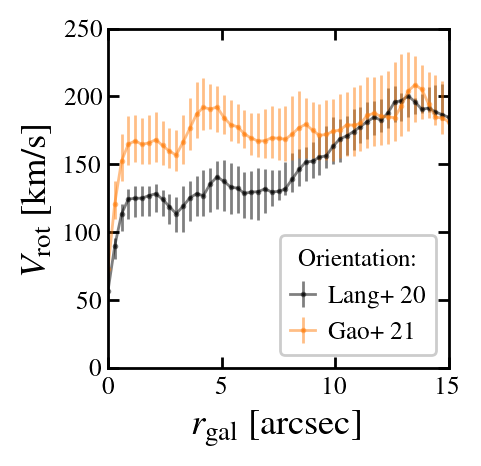}{0.31\textwidth}{}
    \fig{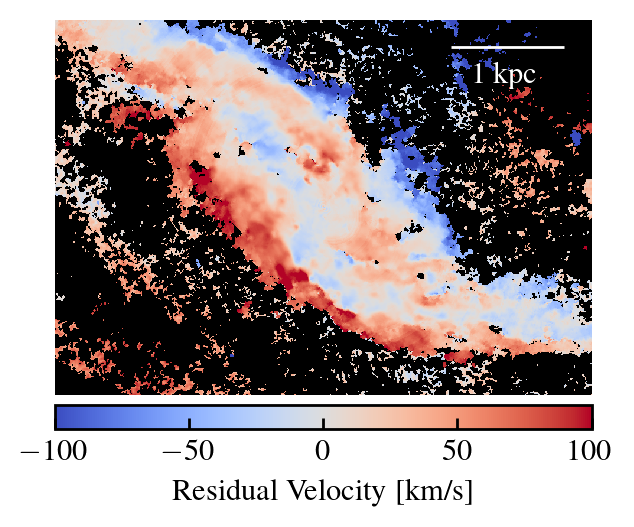}{0.36\textwidth}{}
    \fig{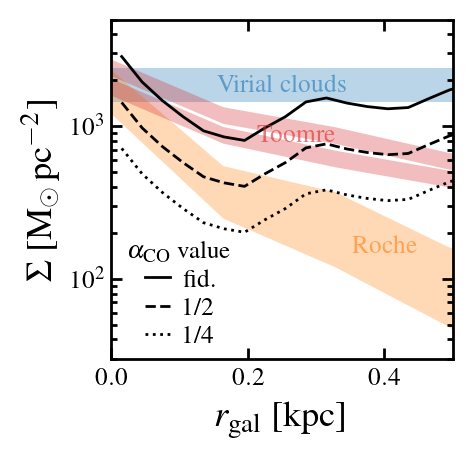}{0.305\textwidth}{}
    }
    \vspace{-2.5\baselineskip}
    \caption{
    Refined molecular gas rotation curves using the 0.3$\arcsec$ CO velocity field are shown using the orientation from \citet[][$\rm PA=201^{\circ}, inclination=55^{\circ}$, black curve]{lang2020} and of \citet[][$\rm PA=220^{\circ}, inclination=40^{\circ}$, orange curve]{Gao2021} (\textit{left}). The resulting residual velocity field when removing a circularly rotating disk (based on the inferred rotation curve using the \citet[][]{Gao2021} orientation) clearly reveals strong streaming (i.e. non-circular) motions at the location of the gas lanes (\textit{middle}). Radial profile of the median molecular gas surface density (nominal curve: black solid line, assuming a 2--4$\times$ lower CO-to-H$_2$ conversion factor: dashed and dotted lines, respectively) is compared to a number of stability criteria:  virialized clouds (blue), Toomre critical density (red), and Roche density (orange) (\textit{right}). 
    The gas in the inner disk ($\rm R_{gal} \leq 5\arcsec$) is basically stable against gravitational collapse for all criteria considered (see text for details).
}
    \label{fig:rot} 
\end{figure*}

We apply {\sc ScousePy} \citep{Henshaw2016,Henshaw2019} to the ALMA 0.3\arcsec\ CO(2-1) data cube to decompose the emission lines into individual Gaussian components (Fig.\,\ref{fig:scouse}; see Appendix \S\,\ref{appendix:scouse}). 
Following quality control, $\sim93\%$ of all spectra contained within the masked region have model solutions. The majority of the line emission in the inner 5\,kpc is well fitted with a single component, with multi-component models required for $\sim31\%$ of the spectra. On average, $\sim1.4$ components are fit per spectrum (or $\sim2.1$ in the spectra where multi-component models are needed). The exceptions to this are the brighter bar lane regions where almost always two or more components are required. In the regions with high effective widths (EWs), for example in the southern lane (top left panel Fig.\,\ref{fig:ngc1365_co}), four, five, or even six component models are sometimes necessary to describe the data (top right panel Fig.\,\ref{fig:ngc1365_co}). These individual components are distinct features in the line profiles, typically spread over a velocity range of $\approx 125\,\mathrm{km\,s}^{-1}$, with velocity separations of $\rm \delta v \approx 41\,\mathrm{km\,s}^{-1}$ (cf. the mean measured dispersion of these components of $\approx17\,\mathrm{km\,s}^{-1}$). The relative motion between these distinct peaks can help to drive an increase in the measured EW at these locations and this velocity separation suggests significant inter-cloud motion.

Although the spread in measured quantities is large, the distributions of the inferred integrated line intensities and velocity dispersions are reasonably similar for single and multiple component fits. As can be seen in Fig.\,\ref{fig:scouse} (bottom left panel), single and multi-component models occupy a similar region in peak temperature--velocity dispersion space. The median peak temperature (amplitude of the Gaussian fit), integrated intensity (integrated Gaussian), and velocity dispersion (sigma of Gaussian fit) are $\rm T_{peak} = 3.6^{+4.1}_{-1.5}\,K$, $\rm I_{CO} = 77.8^{+88.9}_{-42.0}\,K\,km\,s^{-1}$, and $\rm \sigma = 18.6^{+11.1}_{-7.2}\,km\,s^{-1}$, where the upper and lower limits here represent the $\rm 16^{th}$ and $\rm 84^{th}$ percentiles, respectively. We convert the CO intensity of each individual {\sc ScousePy} component to ${\rm H}_{2}$ surface density using the information outlined in \S\,\ref{subsec:alma} and the conversion $\rm R_{21}=1$ outlined in \citet[][]{LEROY2_PHANGSJWST}. The median value for individual components is $\rm \Sigma_{H_2} \sim 800^{+916}_{-432} M_{\odot}pc^{-2}$. Both the median velocity dispersion and the surface density of components are similar to those found in the Central Molecular Zone of the Milky Way \citep[][]{henshaw2022}.

The bottom-right panel of Fig.\,\ref{fig:scouse} shows the three-dimensional, $ppv$ (position-position-velocity), distribution of the individual velocity components (those located within the white dashed box in the top left panel). The color corresponds to the peak brightness temperature of the fitted components in the bar lanes, and we highlight in black all velocity components within an ellipse (with semi-major and semi-minor axes of $2.5\arcsec\times2\arcsec$ or $240\,\mathrm{pc}\times190\,\mathrm{pc}$, respectively; see top left panel) centered on the nucleus (in the plane on the sky). The velocity structure of these highlighted components is consistent with a dynamically cold circularly rotating gas disk, i.e. the components follow a thin 2-dimensional plane in $ppv$ space. 
In contrast, the bar lanes display a comparatively complex velocity structure, with $5-10\,\mathrm{km\,s}^{-1}$ local velocity fluctuations superposed on the large-scale (ordered) non-circular motions. The warp-like appearance in $ppv$-space is consistent with streaming motions. These complex local fluctuations may be driven by a combination of physical mechanisms (e.g., non-circular gas flows, colliding flows), and in-depth analysis of these goes beyond the scope of this initial paper. However, we note that they are qualitatively similar to those detected ubiquitously across all scales in the ISM, both in nearby galaxies and in the Milky Way \citep{Henshaw2020}.

\smallskip
We fit the 0.3\arcsec\ resolution CO velocity field using the software developed by \citet[][]{lang2020} in order to refine the rotation curve of \cite{lang2020} particularly at small galactocentric radii ($R<20\arcsec$) (see Fig.\,\ref{fig:rot} left). 
As the derived rotation curve is less reliable for radii beyond the inner disk that are affected by strong streaming motions, we only show the rotation curves out to $\rm R_{gal}\approx 15\arcsec$.

Two orientations are considered: the orientation used by \citet{Gao2021} \citep[$\rm PA=220^{\circ}; inclination=40^{\circ}$ from][]{Sakamoto2007} and the orientation fitted at 150~pc resolution ($\rm PA=201^{\circ}; inclination=55^{\circ}$) by \citet{lang2020}\footnote{Determination of the position angle and inclination from kinematics is not straightforward for NGC\,1365 due to the presence of strong non-circular motions associated with the bar plus outer spiral arms \citep[e.g. in the CO(1-0) velocity field, see Fig.\,30 of][]{Morokuma-Matsui2022} and NGC\,1365 being the most massive member of a sub-group of the Fornax cluster \citep[][]{loni2021}.}. For reference we note that circular velocity measured in the innermost 30~pc bin in both cases is consistent with the expected motion around a central black hole with mass in the range $\log M_{BH}/M_\odot=7-8$ estimated for this galaxy given its stellar mass and adopting the $\rm M_{BH}$-stellar mass scaling relation measured by \citet{vandenBosch2016}.

For direct comparison to Fig.\,6 of \citet{Gao2021}, the residual velocity field shown (Fig.\,\ref{fig:rot} middle) adopts their orientation parameters. The pattern of residuals in the inner disk is consistent with a rotating disk, but shows evidence that either the adopted inclination is incorrect or, alternatively, that the gas is still moving on elliptical orbits associated with the bar potential. Towards larger galactocentric radii, strong non-circular, streaming motions associated with the bar lanes are well separated from the inner disk and are consistent with those shown by \citet{Gao2021}.
A full kinematic analysis combining information across the full gas disk from atomic and molecular gas data would be able to address these open points, but it is beyond the scope of this paper.

We derive dynamical timescales of $\rm t_{dyn}(R=5\arcsec \approx 475\,pc)\sim 15-30\,Myr$ and $\rm t_{dyn}(R=10\arcsec \approx 950\,pc)\sim 36-40\,Myr$ from the orbital period at these radii. These short timescales imply that the morphology seen in cold gas, dust, and star formation (tracers) can evolve quickly and they are comparable to the timescales inferred for the gas cycling of molecular clouds \citep{kruijssen2019,schinnerer2019,chevance2020,kim2022,pan2022,ward2022}.

To assess the stability of the inner gas disk against gravitational collapse, we compare the molecular gas surface density to three reference critical densities: the Roche density \citep[][]{tan2000} for tidal stability on the beam scale (30~pc)
\begin{equation}
\Sigma_{tidal}=\frac{r_c}{R}\left(\frac{V_c^2}{R}-\frac{dV_c^2}{dR}\right),
\end{equation}
the Toomre critical density \citep[for stability with Toomre parameter $Q=1$;][]{toomre1964}
\begin{equation}
\Sigma_T=\frac{\sigma\kappa}{\pi G}
\end{equation}
in terms of the radial epicyclic frequency $\kappa^2=2\Omega^2(\beta-1)$ where $\Omega=V_c/R$ and $\beta=d ln V_c/d ln R$,
and the surface density needed for uniform density spherical clouds with sizes $r_c\approx 30\,pc$ (one beam) to reach virial equilibrium \citep[with virial parameter $\rm \alpha_{vir}$=1;][]{bertoldi1992}, 
\begin{equation}
\Sigma_{vir}=\frac{5\sigma^2}{\pi G r_c}.
\end{equation}

To estimate the Toomre and Roche critical densities, which both depend on the rotation curve, we have used the rotational velocities newly fitted at 0.3" adopting the \citet{Gao2021,Sakamoto2007} orientation as well as the rotational velocities fitted adopting the \citet{lang2020} orientation. 
Radial derivatives of each rotation curve are calculated with windowing to penalize fluctuations that are inherited with the discrete nature of the measurements.  
The uncertainty associated with the choice of rotation curve denotes a spread in the estimated values at a given galactocentric radius.

For the Toomre and virial reference surface densities we consider a range of 30\,pc scale velocity dispersions $\sigma=14-18$ km s$^{-1}$ based on the typical single-component dispersions fitted by {\sc ScousePy} in the region. This range introduces a spread in the virial density at a given radius. The upper and lower values of the adopted $\sigma$ define two estimates of the Toomre critical density, each otherwise dominated by rotation curve uncertainty.  

As can be seen in Fig.\,\ref{fig:rot} (\textit{right}), the gas surface density lies consistently below the reference stability thresholds inside 300\,pc, epecially when the conversion factor is $\rm 2-4\times$ lower (see \S\,\ref{subsec:alma}). To identify the exact mechanism responsible for preventing the molecular gas from collapsing requires further in-depth analysis, e.g. taking into account the gas structure. 

\section{Discussion}
\label{sec:discussion}

Using new $\sim$30\,pc resolution observations of the molecular gas in CO(2-1) from ALMA and dust emission from JWST/MIRI, we establish a picture of the star formation process in the past $\sim$10\,Myr (as traced by the young star clusters) in the inner $\sim$5\,kpc of NGC\,1365's disks. One of our key findings is evidence for rapid time evolution in the gas distribution and its impact on the present-day distribution of star formation: We see regions undergoing intense massive star formation and some regions that do not appear to be forming stars at all. Before summarizing the main properties of these regions in $\S$ \ref{subsec:SFring}, we highlight relevant aspects of the dynamics of the bar environment, assembled from the CO observations analyzed in this work (\S\,\ref{subsec:flows}). The star formation appears sensitive to the current molecular gas distribution that is expected to rapidly evolve in the central regions of the bar given orbital dynamics and strong inflows \citep[see also][]{Sakamoto2007}. In \S\,\ref{subsec:simulations} we leverage hydrodynamical simulations of a galaxy with properties similar to NGC\,1365 to support this view, in contrast to other work suggesting that (at least partially) the AGN outflow piles up the gas and induces star formation \citep[][also discussed in \S\,\ref{subsec:starburst}]{Gao2021}. 

\subsection{Gas Flows and Turbulent Motions in the Bar}
\label{subsec:flows}

We can gain insight into the nature of star formation in the bar by considering the observed molecular gas properties in the context of bar dynamics. Gas flows along so-called $x_1$ bar orbits and along the set of perpendicular $x_2$ orbits which form a ring or ellipse (see Fig.\,\ref{fig:ngc1365_overview}) evident in the F770W map. The transition from $x_1$ orbits to the $x_2$ orbits leads to the development of the bar lane shocks along the bar leading edge (the northern and southern bar lanes), where gas is funneled towards smaller radii \citep{Athanassoula1992,Sormani2015}.  

The fueling of the central ring-like structure by gas from larger radii appears to be at least partially responsible for the complex kinematics in the region as revealed by the {\sc ScousePy} decomposition. Gas arriving at the ring shares the same high level of complexity present in the northern and southern bar lane arms (described more below; see Fig.\,\ref{fig:scouse}). In addition, the ongoing massive star formation in the ring-like structure is also injecting a large amount of energy via stellar feedback. 
The disappearance of the complex velocity structure towards smaller galactocentric radii suggests that the sources contributing to the complex velocity structure are changing towards smaller radii (see \S\,\ref{sec:conclusion}). 

Further out (beyond the inner $R_{\rm gal} \geq 12\arcsec$) in the northern and southern bar lanes that run along the large-scale bar \citep{Leroy2021-Survey,Egusa2022}, overlapping gas streamlines lead to more complex gas kinematics, often with multiple peaks observed along the same line of sight. Ongoing massive star formation, meanwhile, is distributed in an asymmetric fashion along the bar lanes with many cluster candidates associated along the full length of the northern lane while the southern lane lacks clusters at larger radii despite fairly high integrated CO intensities. There is a broad age trend when going from large to small galacto-centric radii along the bar lanes.

In the northern lane (see Fig.\ref{fig:co_sf}), the highest integrated CO intensity coincides with the highest attenuation and the brightest MIR clusters at large radii whereas low CO intensity is co-located with the brightest observed H$\alpha$ emission and a clustering of young clusters at smaller radii. Further molecular gas properties in that region are consistent with more diffuse, warm gas relative to the rest of the ring \citep{LIU_PHANGSJWST}, this is consistent with this region having experienced star formation for a longer amount of time. 
For the southern lane, the lack of star formation in the outer part of the lane together with the high attenuation implies that this region is less far along in the star formation process than the inner part where many clusters are seen and attenuation is low. The time difference between the two bar lanes is about 10\,Myr when considering only the dynamical time. It is interesting that while a rough time sequence is evident along both bar lanes, it is not exactly mirrored as the southern lane contains molecular gas before the onset of massive star formation.

\subsection{Gas, Dust, and Star Formation at 30\,pc Resolution}
\label{subsec:SFring}

\begin{figure*}
    \centering
    \includegraphics{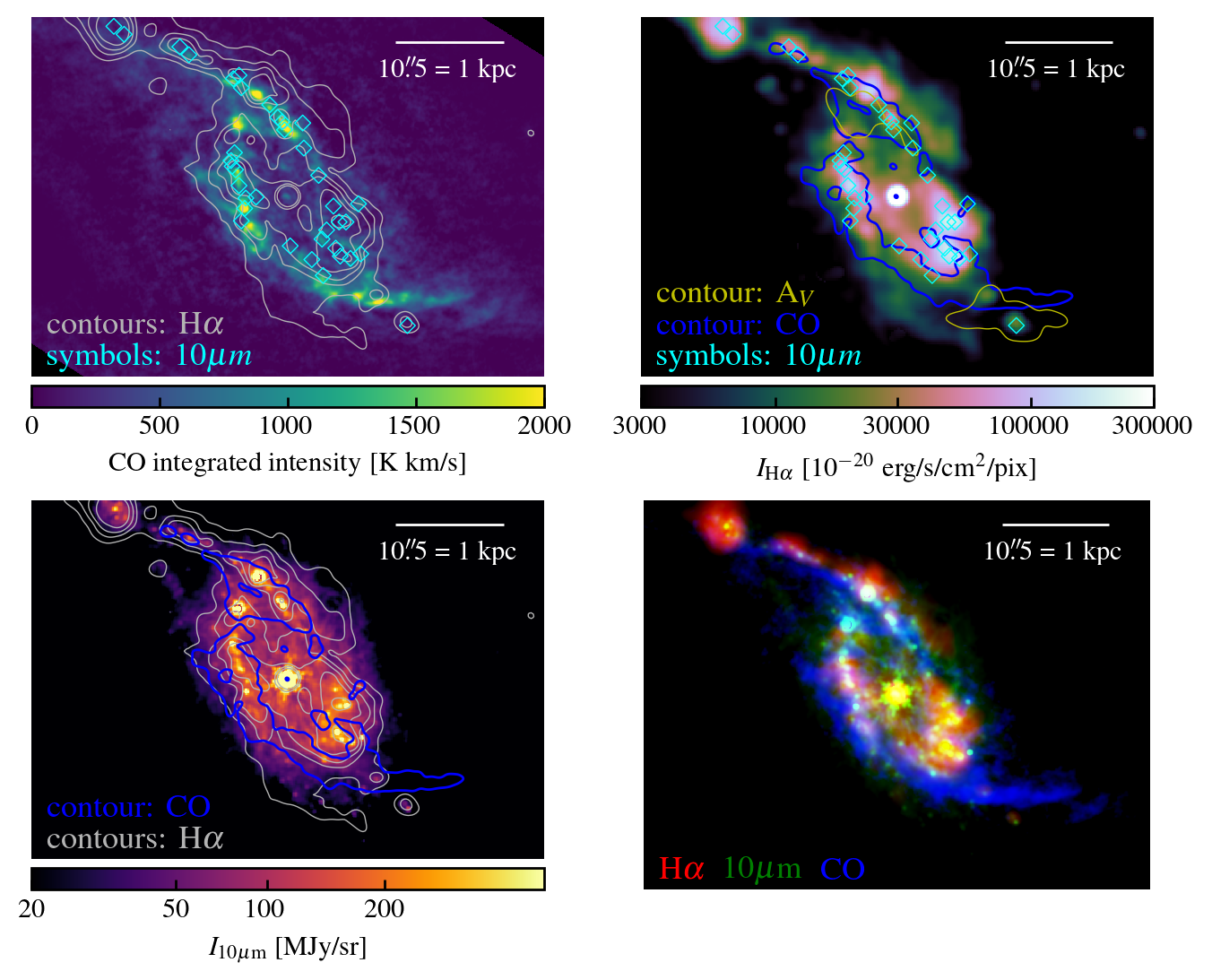}
    \caption{Comparison of the distribution of the molecular gas and sites of ongoing star formation in the central $\sim$5\,kpc of NGC\,1365 in four different representations (North is up, East to the left). The ALMA CO(2-1) integrated intensity map (in color: \textit{top left}, in blue: \textit{bottom right}, as blue contour: remaining panels) basically traces the immediate gas reservoir for star formation. The JWST/MIRI 10\,$\mu$m continuum map (in color: \textit{bottom left}, in green: \textit{bottom right}) where the bright compact regions (cyan symbols, \textit{top panels}) are likely tracing young ($\rm \leq 10\,Myr$) massive ($\rm \approx 10^6\,M_{\odot}$) star-forming regions. The PHANGS-MUSE H$\alpha$ emission line map \citep{Emsellem2022} at 1.15\arcsec\ resolution (in color: \textit{top right}, in red: \textit{bottom right}, as grey contours in remaining panels) reveals the location of bright HII region complexes that are ionized by massive O stars. Regions of high attenuation \citep[$\rm A_V\gtrsim\,4\,mag$;][]{Belfiore2022} are indicated by a yellow contour.
    }
    \label{fig:co_sf} 
\end{figure*}

\begin{figure*}
    \centering
    \includegraphics[width=1.0\textwidth]{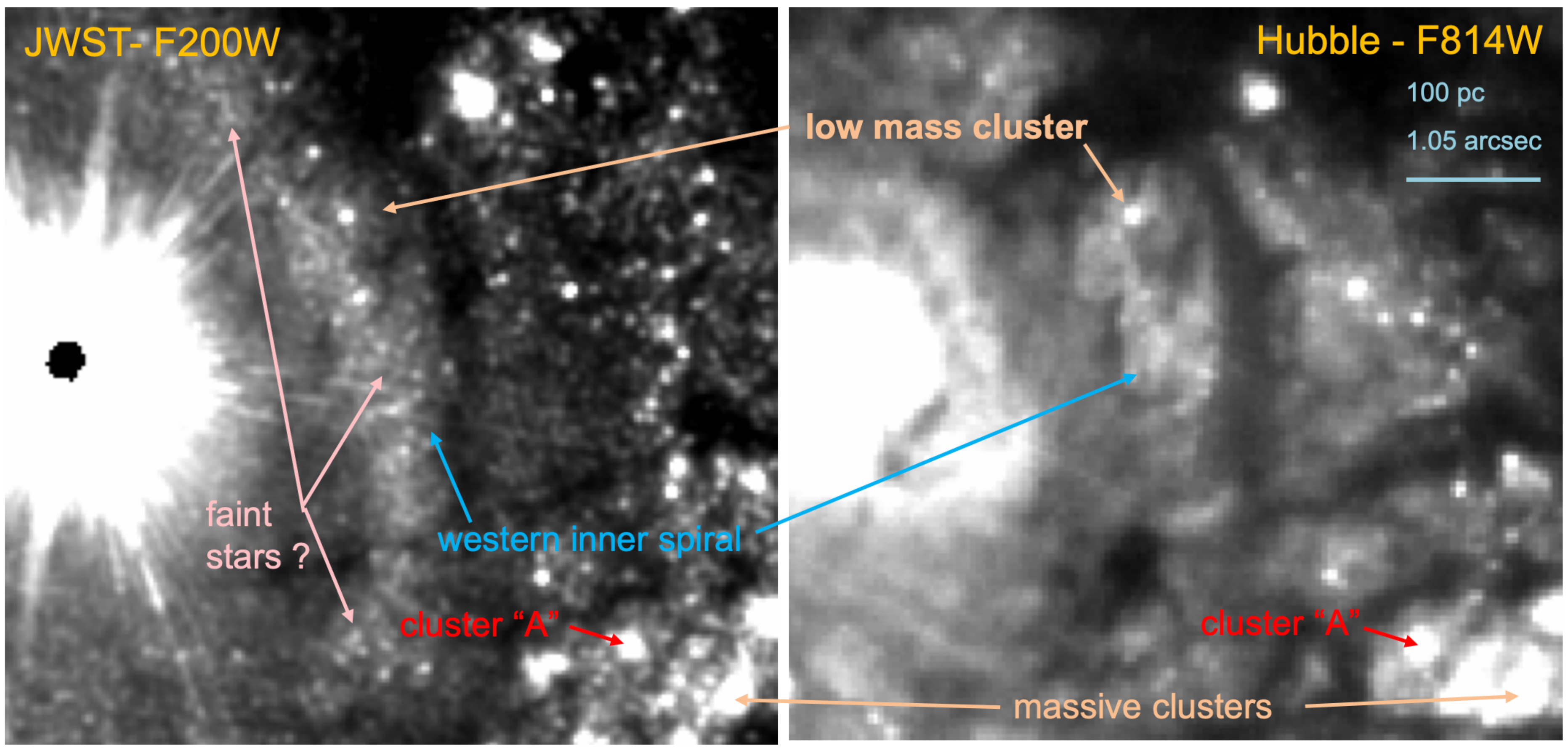}
    \caption{Close-up look (5\arcsec$\times$5\arcsec) of stellar populations in the inner disk (North is up, East to the left). The high sensitivity and resolution of the JWST/NIRCAM F200W imaging from PHANGS-JWST \citep[\textit{left};][]{LEE_PHANGSJWST} reveals more structure in the region where undisturbed circular rotation gas has been found compared to what can be seen in the HST F814W map \citep[\textit{right};][Whitmore et al. subm.]{Lee2022,WHITMORE_PHANGSJWST}.
    Different regions that are discussed in the text are annotated. The bright (saturated) source to the east is the AGN at the center of NGC\,1365.
    }
    \label{fig:center} 
\end{figure*}

The inner 5\,kpc of NGC\,1365 can be divided in two broad regimes given its molecular gas and star formation properties (see Fig.\,\ref{fig:co_sf}): those exhibiting no/low star formation (central smooth disk, outer southern bar lane) and those exhibiting a high abundance of young massive clusters (massive star-forming region southwest of the center, string-like star formation northeast of the center).  

We expect these regions to be common to the CMZs of other barred galaxies, although their occurrence could depend on the exact geometry of the underlying bar orbits, the distribution of gas along the bar, and the time when the system is observed.
We thus give an overview of the basic properties of these regions, and comment on how they support our view of rapid evolution in the gas and star formation distributions in the bar region.  We also comment on the nature of the central starburst.

\subsubsection{Low star formation regions}

The absence of massive young star clusters is evident for the inner 1\,kpc as well as the outer part of the southern molecular bar lane.

\smallskip
\textit{Central smooth disk --}
We find no evidence of ongoing massive star (cluster) formation inside a galacto-centric radius of $\rm R_{gal}\approx5\arcsec\ (475\,pc)$ where a dynamically cold regularly rotating molecular gas disk has been identified. 
Figure \ref{fig:center} shows an image of the western inner ``spiral-like" feature near the nucleus of NGC 1365 \citep[see also][]{WHITMORE_PHANGSJWST}. We find that while it breaks down into hundreds of individual point-like objects in the F200W map from PHANGS-JWST \citep{LEE_PHANGSJWST}, it remains patchy and nebulous in the F814W image from PHANGS-HST \citep[e.g.][Whitmore et al. subm.]{Lee2022}. There is only one object along the feature that is bright enough to be a regular, low-mass clusters (i.e., $\rm \sim10^5\,M_{\odot}$; B. Whitmore, priv. comm.). We speculate that either massive clusters, like those that exist just a few hundred parsecs away in great number, cannot form in this environment, or perhaps more likely, the massive clusters formed at larger radii are destroyed by tidal forces, leaving the debris of large numbers of individual stars in their wake in a visibly smooth looking pattern \citep[similar to the scenario proposed for the massive star clusters in the Milky Way CMZ by][]{habibi2014}. We note that the hundreds of faint, crowded, point-like objects along the western inner spiral have F200W fluxes similar to red supergiant stars with ages around 10\,Myr seen in the outer parts of the galaxy, but since they are not resolved in the visible, or at longer wavelengths due to available spatial resolution, it is difficult to determine their color and thus true nature. 

The mean integrated CO(2-1) intensity in the inner 1\,kpc is $\rm \langle I_{CO(2-1)}\rangle \approx 300\,K\,km\,s^{-1}$ which corresponds to gas surface densities of $\rm \Sigma_{H_2}\approx 1300\,M_{\odot}pc^{-2}$ assuming a Galactic conversion factor and $R_{21}=1$ (see \S\,\ref{subsec:alma}). Lowering the conversion factor by 2-4$\times$ \citep{Teng2022} still results in sufficiently high surface densities for star formation to happen. However, rough estimates of critical densities for stability (see \S\,\ref{subsec:scouse}) suggest that even at these high gas surface densities the gas is stabilized against gravitational collapse.
Overall, these properties are similar to the central molecular disk in our own Milky Way, where it has been suggested that high shear can dissolve and disrupt clouds \citep[e.g.\ ][]{hatchfield2021} and act to counter gas self-gravity \citep{meidt2018,li2020}. The central region is also reminiscent of the situation in early type galaxies (ETGs), where strong shear is able to oppose self-gravity \citep{liu2021} and prevent the fragmentation of the gas disk \citep{gensior2020}, thus suppressing the formation of massive stars \citep[][]{davis2022}. This would mean that star formation in the inner 1\,kpc is suppressed over long timescales.

\smallskip
\textit{Outer southern bar lane --}
There is much reduced ongoing massive star formation in the outer southern bar lane (only two 21\,$\mu$m selected regions, but no 10\,$\mu$m identified cluster candidates and only one H$\alpha$ emission peak being a HII region; see Fig.\,\ref{fig:co_sf}). The apparent lack of corresponding 7.7\,$\mu$m PAH emission (see Fig.\,\ref{fig:jwst} left) is puzzling. We speculate that this could be related to the lack of heating sources (cf. outer part of northern bar lane), as there is only one H$\alpha$ peak which has E(B-V)$\approx$1.7\,mag (or $\rm A_V\approx5.3\,mag$). In contrast to the central disk discussed above, this lack of star formation points to a timing issue (see also \S\,\ref{subsec:simulations}). While we observe the highest CO velocity dispersion (coincident with high integrated CO intensity but only average peak temperatures; see Fig.\,\ref{fig:ngc1365_co}) in this region, the {\sc ScousePy} decomposition shows that the high dispersion arises due to multiple line components that have peak temperatures and line widths consistent with the remaining molecular gas (see Fig.\,\ref{fig:scouse}). Thus we can rule out enhanced turbulence as the cause for the lack of star formation. Taken together this suggests that this region is the (relatively speaking) youngest region in the inner part where star formation has not yet turned on. Hydro-dynamical simulations of barred galaxies show a rapid evolution in the gas distribution in their CMZ \citep[][see also \S\,\ref{subsec:simulations}]{tress2020,Sormani2020}.

The absence of ongoing massive star formation has different origins in the two regions discussed above. While in the central disk, there are factors that can genuinely suppress star formation, the situation for the outer southern bar lane could be just a timing issue given that the molecular gas properties are not different from the northern bar lane that hosts abundant young star clusters (see \S\,\ref{subsec:simulations} and \S\,\ref{sec:conclusion}). Thus one needs to be cautious when interpreting such results, especially since there is abundant massive cluster formation in the remaining parts of the CMZ.

\subsubsection{High star formation regions}

It is notable that all 37 cluster candidates identified at 10\,$\mu$m coincide with regions of notable H$\alpha$ emission (above $\rm 10^{16}\,erg\,s^{-1}\,cm^{-2}\,spaxel^{-1}$), suggesting that H$\alpha$ emission is a very good indicator of the location of active star formation and almost no sites of highly embedded star formation are missed even in the gas-rich southern lane (Fig.\,\ref{fig:co_sf} top right). This spatial coincidence is not surprising given the high coincidence ($>$ 90\%) of 21\,$\mu$m-selected ISM sources with PHANGS-MUSE HII regions \citep{HASSANI_PHANGSJWST}, the good correlation between CO, H$\alpha$, and MIRI (F770W, F1000W, F1130W, F2100W) emission using $\sim$100\,pc-sized pixels \citep{LEROY1_PHANGSJWST} and the significant overlap ($\sim$70\%) between 24$\mu$m (21$\mu$m) and H$\alpha$ emitting time-scales \citep[][this issue]{kim2021,KIM_PHANGSJWST}. The distribution of young clusters and massive star formation can be divided in two groups.

\smallskip
\textit{Massive star formation southwest of the center --}
The relatively smooth distribution of the brightest 7.7\,$\mu$m PAH emission (see Fig.\,\ref{fig:jwst}) suggests cold gas has been accumulating at the CMZ for some time.
There are two interesting regions along this PAH `ring'. First, there is a clustering of ten ($\sim$30\%) young massive star cluster candidates identified at 10\,$\mu$m across radii of 5-10\arcsec\ at the inner end of the northern bar lane, i.e. southwest of the nucleus just north of the outer southern molecular gas lane. This region exhibits only faint CO emission and is the location of the brightest observed H$\alpha$ emission indicating the presence of massive HII region complexes (see Fig.\,\ref{fig:co_sf}, top row). \cite{LIU_PHANGSJWST} study the molecular gas excitation in the central 5\,kpc at $\sim$330\,pc resolution and find evidence for low-density warm molecular gas as being the dominant phase in this southwest region. Taken together this suggests that the star formation in this region has significantly impacted the surrounding cold molecular gas and might be -- relatively speaking -- most evolved in time.

\smallskip
\textit{String-like star formation northeast of the center --}
Along both bar lanes northeast of the nucleus, cluster candidates follow closely the bright CO ridges, but they do not always coincide with peaks in CO emission, and 2/3 of all 10\,$\mu$m selected candidates are located here (Fig.\,\ref{fig:co_sf} top left, see also Fig.\,\ref{fig:jwst} right). Observed H$\alpha$ emission from this area is fainter than in the region southwest of the nucleus discussed above, however, attenuation inferred from the Balmer lines (H$\alpha$,H$\beta$) reaches values above $\rm A_V=4\,mag$ where both bar lanes appear to connect and the brightest (saturated) 10\,$\mu$m clusters are. The inferred SFR surface density map, i.e. H$\alpha$ emission corrected for attenuation, shows clearly that massive star formation ($\rm \Sigma_{SFR} > 1\,M_{\odot}yr^{-1}kpc^{-2}$) is  confined to the CO ridges and the southwest region. Interestingly, the inner end of the southern bar lanes shows much less attenuation ($\rm A_V\leq 2.5\,mag$). Given the close morphological relation between the CO ridge lines and the distribution of the massive cluster candidates we speculate that the observed clusters are still fairly close to their formation sites. 

It is interesting to note that star cluster candidates are found along the outer northern bar lane, i.e. outside the PAH `ring', implying that conditions for massive star formation to occur are met in the northern lane already outside the inner structure. \cite{WHITMORE_PHANGSJWST} suggest that the large populations of somewhat older star clusters, with ages between 20 and 400\,Myr, that are found slightly outside the northern bar lane, originally formed along the bar at larger galacto-centric radii. Due to the dynamical decoupling of these stellar clusters from the gas flow along the bar (and onto the ring), these populations are now residing in the so-called overshoot region (see also Fig.\,\ref{fig:ngc1365_overview} bottom right). It is conceivable that the currently young clusters could be the precursors of such a population of clusters in a few 10\,Myr.

\subsubsection{On the origin of the starburst and role of AGN outflow} 
\label{subsec:starburst}

\cite{boeker2008} proposed two scenarios on how star formation might proceed within a starburst: (a) pearls-on-a-string scenario where star formation is preferentially triggered at or close to the location where the gas lanes along the large-scale bar connect to the inner gas structure, and (b) pop-corn scenario where star formation occurs stochastically within the ring with no age trend. The rough age trend in overall star formation present along the bar lanes -- from star-forming molecular gas to star formation heated molecular gas in the northern lane, from non-star-forming molecular gas to star-forming molecular gas in the southern one. This points toward an evolution of the star formation process along the gas lanes as expected for the pearls-on-a-string scenario. However, there is no preferred triggering point, rather star formation starts already before the gas reaches the so-called contact point (see northern lane) or well after it entered the inner structure (see southern lane). The dynamical time of 10\,Myr for one gas lane of the ring is sufficiently long so that age trends
among individual star clusters can be used to more robustly discriminate between the two scenarios when more accurate age estimates based on SED fitting become available in the future.

\smallskip
NGC\,1365 hosts a large ($r>2$\,kpc) biconical outflow in ionized gas that is likely driven by its central AGN \citep[e.g.,][]{jorsater1984,edmunds1988,hjelm1996,Lindblad1996, Lindblad1999,lena2016, Venturi2018}. 
\citet[][]{Gao2021} suggested that this outflow is impacting the molecular gas disk based on the analysis of $\sim$2\arcsec\ CO(1-0) imaging and affecting the star formation activity in the starburst ring. Our results do not favour such a scenario.

The inner ($\rm R_{gal}<5\arcsec$) dynamically cold unperturbed central gas disk is consistent with circular rotation. The {\sc ScousePy} decomposition reveals no evidence for second line components which could be associated with the outflow impacting the molecular disk \citep[for an example, see, e.g., M\,51;][]{querejeta2016} which is consistent with \citet[][]{combes2019} who find no evidence for a molecular outflow in their $\sim$0.1\arcsec\ resolution CO(3-2) data.
The on-sky orientation of the outflowing ionized gas is in a fan-like geometry to the southeast and northwest of the nucleus \citep[e.g., see Fig.\,6 in][]{Venturi2018}. This would imply that molecular gas in the bar lanes discussed here is not impacted in their outer parts. We see no evidence for a change in the characteristics of the complex gas structure revealed by {\sc ScousePy} in Fig.\,\ref{fig:scouse} along the gas lanes, i.e. the complexity is similarly high along the full range of gas lanes probed. Further,
there is no variation in the distribution of the 7.7\,$\mu$m PAH emission.
Lastly, \citet{LIU_PHANGSJWST} find the CO excitation and CI/CO line ratio to be consistent with star-forming regions in the disk.

Taking all points together makes it highly unlikely that the AGN is impacting the molecular gas in the central region and implies a scenario where the outflow is more aligned with the rotation axis, so that it is not even grazing the molecular gas disk. This is consistent with the 3D cone model of \citet{hjelm1996} based on fits to ionized gas emission lines. Their inferred cone axis is within $5^\circ$ of the galaxy's rotation axis, so that the cone is pointing outward the galaxy disk with an opening angle of 100$^\circ$.

\subsection{Evolution of Gas and Star Formation Distribution}
\label{subsec:simulations}

\begin{figure*}
    \centering
    \includegraphics[width=\textwidth]{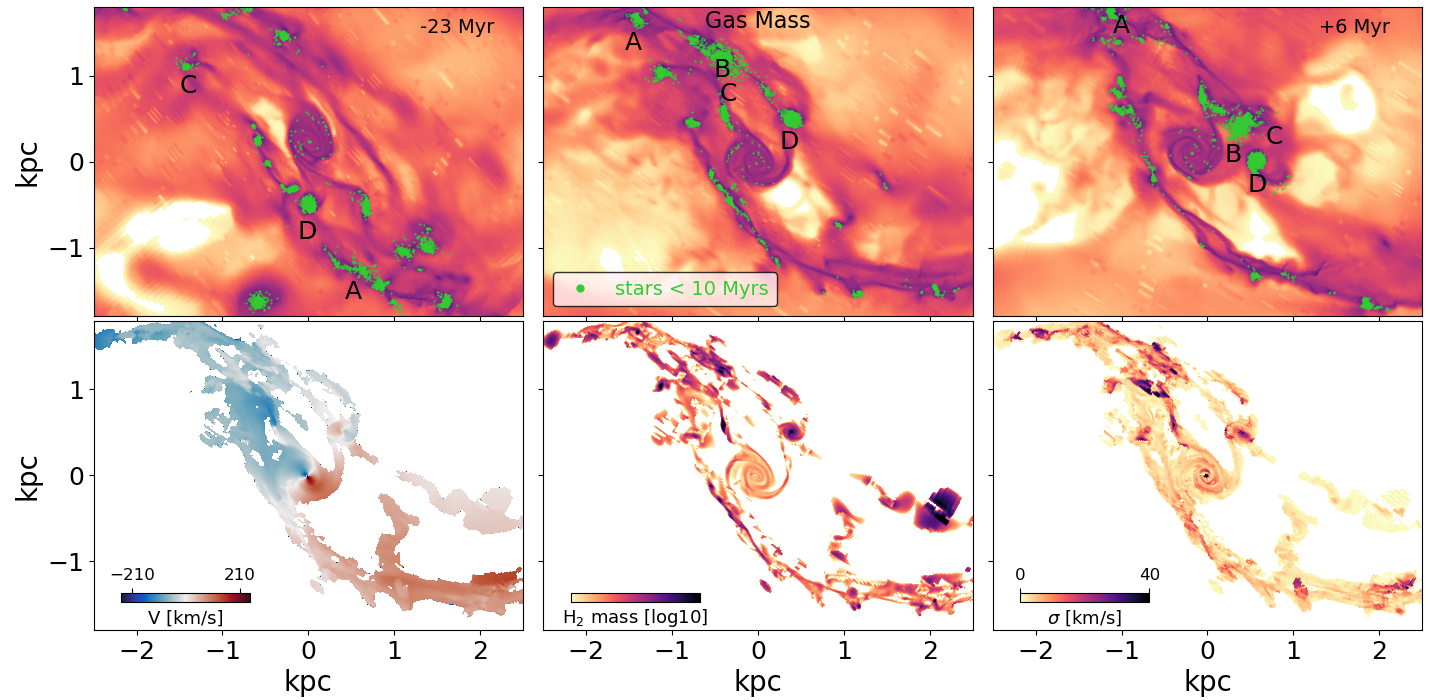}
    \caption{Projected gas density and kinematic maps from a hydro-dynamical simulation of a barred galaxy with NGC\,1365-like properties. The selected snapshot broadly matches the properties of the gas in NGC\,1365's inner disk. The central $\rm 5\times3.5\,kpc^2$ with the large-scale galactic bar position and disk orientation on-sky matched to NGC\,1365's properties. \textit{Top row:} Distribution of the gas mass (color) and stars formed in the past 10\,Myr (green points) for the selected snapshot (\textit{middle}). Comparison to two epochs showing the distributions 23\,Myr earlier (\textit{left}) and 6\,Myr later (\textit{right}) reveals the fast evolution of the morphology in just $\sim$30\,Myr which roughly corresponds to the dynamical time at a galactocentric radius of $\rm R_{gal}=1$~kpc. Four prominent star clusters are labeled in the panels.
    \textit{Bottom row:} Properties of the molecular H$_2$ gas: gas velocity field (\textit{left}), $\rm H_2$ gas mass distribution (\textit{middle}), and $\rm H_2$ gas velocity dispersion
    (\textit{right}).} 
    \label{fig:sim} 
\end{figure*}

In the following we compare the observed properties to a {\sc Ramses} simulation of a galaxy with global properties matched to NGC\,1365 (see \S\,\ref{subsec:ramses}). A snapshot resembling NGC\,1365's inner gas distribution is shown in Fig.\,\ref{fig:sim} (top middle panel). The asymmetric gas morphology consists of a prominent southern gas lane along the bar that extends almost to the northern lane, a shorter northern lane, an additional gas filament located between both lanes northeast of the center, and an inner smooth gas disk plus a couple of compact gas peaks. Similarly, the distribution of young stars with ages of $\rm < 10\,Myr$ is not very symmetric with most star formation having occurred northeast of the nucleus. Star formation has been clustered and the amount of stars formed varies significantly between different sites with more prominent locations to the northeast. Young stars are mostly found along the ridge lines of the gas lanes and are associated with peaks in the gas distribution, but sometimes away from the lanes. This is illustrated by a few small clusters forming in the southern gas lane in possibly Kelvin-Helmholtz instability driven structures \citetext{e.g., \citealt{renaud+2013}; see also \citealt{mandowara2022}}. 
Along the southern lane, young stars are much more sparse at large galacto-centric radii and the few present coincide with gas spurs emanating from the gas arm. In the northern lane, four big clusters of a few $10^6$ and $10^7$~\Msun\ are evident. Inside the smooth central disk (radius of about 300~pc), individual star particles are seen, mostly originating from accreted and stripped clusters. The overall resemblance to the observations is striking as the simulation shows a similar asymmetry in the gas distribution and the location where massive star formation occurs in the `starburst ring' and the additional inner smooth gas disk. We note that no bubbles are apparent in the simulations \citep[cf. the two bubbles at large radii on the northern lane noted by][]{WHITMORE_PHANGSJWST}. 

We compare gas mass and young stars distributions across 30\,Myr which is equivalent to the dynamical time at $\rm R_{gal}=1\,kpc$ in NGC\,1365 (Fig.\,\ref{fig:sim} top row) to gain insights into how transient these morphologies are. After 6\,Myr, the northern bar lane becomes more apparent as the filament between the arms has disappeared (right panel). The change in morphology is even more evident in the young stars where the most prominent location is now just west of the inner gas disk while the overall location of smaller sites has shifted along the bar lanes. Inspection of the snapshots shows that the four prominent
star clusters (labeled in Fig.\,\ref{fig:sim}) are streaming inwards together with an associated gas reservoir.
The comparison to 23\,Myr earlier is more drastic as the gas mass morphology does not even represent gas lanes or a ring-like structure. The vast majority of young stars can be found southwest of the nucleus. Despite these stark changes in the gas mass distribution, star formation is always clustered and mostly associated with gas peaks, something that is certainly significantly influenced by the choice of a star formation threshold and unlike to what is seen in the observations (see further discussion in \S\,\ref{sec:conclusion}). The rapid evolution of the gas and star formation distribution at any given point in time can be traced back to an irregular and asymmetric gas flow to the inner region. This means the underlying orbital structure for the gas flow remains stable \citep[e.g.][]{Sormani2020}, while its population with gas is strongly affected by the fragmentation of the gas disk well outside the inner region studied here.

We show for comparison to the observations the synthetic $\rm H_2$ gas mass distribution, velocity field, and velocity dispersion derived from the matching snapshot (Fig.\,\ref{fig:sim}, bottom row). The asymmetry in gas morphology is more evident now as only the densest regions remain. The inner smooth gas disk is fainter, i.e.\ less massive, compared to the gas in the bar lanes. Similarly, the inner disk is consistent with regular rotation while strong non-circular motion is evident in the gas lanes. This abrupt change from circular rotation to streaming motions is also present in the simulations of the Galactic Center by \cite{tress2020} implying that this is a common feature in modern hydro-dynamical simulations of central regions of barred galaxies. The highest velocity dispersion regions almost all coincide with regions where many young stars are present. The exceptions are two locations in the western end of the southern arm where higher dispersion is seen slightly offset from star-forming sites. 
While the synthetic velocity dispersion shows a similar range in values as the observations, the highest dispersions seen in the observations are associated with regions devoid of ongoing massive star formation. Given the rapid evolution of the central region, it is remarkable that many of the other features are matched.

Comparison between simulations and observations shows that asymmetries in the gas and star formation distribution are a natural outcome of gas flows in galaxy like NGC\,1365 \citep[see also][for the CMZ in the MW]{Sormani2018}, that the exact morphology has a transient nature, and that there is no need to invoke the impact of an AGN outflow/ionization cone to explain the very efficient formation of many massive star clusters.

\section{Toward a Comprehensive Understanding of CMZs}
\label{sec:conclusion}

Compared to the CMZ of the Milky Way, the central starburst ring in NGC\,1365 is enormous, its extent is 9$\times$ larger (while the bar is about twice as long as the MW bar), its SFR is 70$\times$ higher, and its dynamical time is about 6$\times$ longer. While the inferred molecular gas surface densities and line widths are more comparable to those observed in the MW's CMZ \citep[see Tab.\,\ref{tab:ngc1365} and][]{henshaw2022}, they are significantly ($\rm \sim2-3\times$) higher than the values typically observed in galactic disks \citep{sun2022}. At least 37 young ($<$10\,Myr), massive ($\rm >10^6\,M_{\odot}$) star clusters are present in the inner 5\,kpc region, much more than found in the nearby major merger system Antennae \citep{WHITMORE_PHANGSJWST} and the ring contributes about 30\% to the total SFR in NGC\,1365. 

Our in-depth analysis of the molecular gas and star formation properties of the CMZ in NGC\,1365 reveals an inner ($\rm r\le475\,pc$) smooth non-star-forming gas disk similar to those seen in early-type galaxies, a variation in star formation activity along the bar lanes broadly consistent with the pearls-on-a-string scenario (i.e. some location preference for star formation), and no evidence for the AGN impacting the gas disk. This is consistent with \cite{Fazeli2019}, who note tentative evidence for the pearls-on-a-string scenario in the bar lane west of the nucleus based on the analysis of the central 8\arcsec$\times$8\arcsec\ using near-IR IFU data. 

While the dedicated simulation is not meant as a real match of NGC\,1365 structures, and does not include a practical treatment of radiation or dust, it appears consistent with the main observed gas-related features, which can be explained without invoking AGN feedback. This strongly suggests that the gas morphology and ongoing star formation in the inner 5\,kpc of NGC\,1365 is caused by the dynamical configuration imposed by the large-scale stellar bar. This is in line with the semi-global simulations of homogeneous gas inflow by \cite{moon2021} where the gas inflow rate $\rm \dot{M}_{in}$ controls the SFR in the ring and a close relation of $\rm SFR \approx 0.8\,\dot{M}_{in}$ emerges. The follow-up work studying varying gas inflow rates links the variation in SFR mostly to changes in the inflow rate with only a moderate impact by stellar feedback from supernovae and shows that only large asymmetric inflow rates can create a lopsided star formation distribution \citep{moon2022}. The result of our simulation is also fully consistent with results from simulations of gas flow in a Milky Way-like barred potential of \cite{Sormani2018}, which show that the asymmetric and transient gas morphology in the CMZ is induced by large-scale gas flow driven by the bar. These simulations also showed that gas flow in a barred potential is essentially always intrinsically unstable, and is bound to develop asymmetric and transient morphologies even in the absence of stellar feedback.

The predicted fast evolution and transient molecular gas pattern for NGC\,1365 from the {\sc Ramses} simulations can explain the lack of a strong pattern in the age distribution. Note that \cite{Seo2013} conclude from their simulations that an azimuthal age gradient in star clusters can only develop for low gas inflow rates ($\rm \lesssim 1\,M_{\odot}yr^{-1}$ for their model setup) when the gas entering the ring by switching from the $x_1$- to the $x_2$-orbits at the contact points has enough time to collapse and form stars. For higher inflow rates, too much gas enters the $x_2$-orbits in order to be directly converted into star clusters and, thus, can continue along the $x_2$-orbits leading to a stochastic age distribution. As \cite{Elmegreen2009} estimate a molecular gas mass inflow rate of $\rm \sim50\,M_{\odot}yr^{-1}$ (corrected to our assumed distance, but keeping their 4$\times$ lower conversion factor from \citealt{Sakamoto2007}), a stochastic age distribution could be expected. 

The lack of a strong pattern in the age distribution is also supported by the simulations of \cite{Sormani2020}. These authors compared the instantaneous and time-averaged (over periods of $\sim 20 \rm Myr$)  distributions of $\rm H_2$ surface density and the youngest stars in their simulation of the Milky Way's CMZ. They find that the instantaneous gas surface density is clumpy and does not show any evident patterns, while the time-averaged distribution is relatively smooth, with young stars that show a slight preference to form downstream from the apocenter, kind of consistent with the pearls-on-a-string scenario. However, these simulations show that the pearls-on-a-string scenario only becomes evident after taking the time-average, and it is very hard if not impossible to detect from a single snapshot as it is available for observations. 

Based on simulations with and without gas self-gravity and stellar feedback, \cite{tress2020} conclude that both are required to move gas from the CMZ towards smaller galactocentric radii inside the ring where it can form a circum-nuclear disk (CND). Although their simulations were set up to mimick the situation of the CMZ in the Milky Way, we speculate that the smooth central disk seen in NGC\,1365 is fed by a similar mechanism as the {\sc Ramses} simulations also include stellar feedback.

The molecular gas morphology in NGC\,1365's central 5\,kpc is less symmetric than that seen in most other circumnuclear gas rings from barred PHANGS-ALMA galaxies observed at $\sim$ 1\arcsec\ (S. Stuber et al. in prep). This could imply that other barred galaxies have less variation in the morphology of their central gas reservoir. Clearly more high-resolution observations are required to build up the necessary statistics to address this question. 

\section{Summary and Conclusion}
\label{sec:summary}

By combining 0.3\arcsec (30\,pc) resolution PHANGS-JWST MIRI and NIRCAM imaging and archival ALMA CO(2-1) mapping with lower angular resolution MUSE data of the central 5\,kpc of the nearby barred spiral galaxy NGC\,1365, we investigate how star formation proceeds in its wide central starburst ring. The MIRI F770W and F1000W filters reveal the distribution of the neutral gas traced via PAH emission and the location of (embedded) young massive star clusters, respectively, while the ALMA CO data provides access to the molecular gas properties in this region. Together these datasets provide an unprecedented view of a CMZ undergoing a period of intense star formation. Comparison to hydro-dynamical {\sc Ramses} simulations of a galaxy with global properties matched to those of NGC\,1365 lends insights on the physics underlying the star formation in this distinct structure and CMZs as such. 
In particular we find:

\begin{itemize}

\item The central molecular gas reservoir is resolved into bright clumpy bar lanes that surround a fainter smooth inner $\rm R_{gal} \sim 5\arcsec (\approx 475\,pc)$ gas disk. This rotating disk is dynamically cold, undergoing circular rotation, and devoid of young massive star clusters. Comparison to stability criteria suggests that this disk is stable against gravitational collapse -- similar to what is observed in the central molecular gas disks of early type galaxies. As a similar structure develops in simulations via gas inflow caused by stellar supernova feedback (and gas self-gravity), such circumnuclear disks (CNDs) should be a common feature, with actual sizes depending on the exact properties of the large-scale bar and the overall molecular gas reservoir available.

\item The star formation distribution is lopsided. Most young massive star cluster candidates are located along the molecular bar lanes. Some can even be found outside the ring structure implying that massive clusters are already forming along the bar, i.e. outside the region dominated by $x_2$-orbits. In the outer southern bar lane, almost no clusters exist, although abundant molecular gas with properties similar to the rest of the bar lanes is present. We speculate that this region is observed just before the onset of star formation.

\item The gas kinematics reveal streaming, i.e., non-circular in-plane, motions and multiple line components in the bar lanes. {\sc ScousePy} decomposition finds average line widths of $\rm \langle \sigma_{CO,scouse}\rangle \approx 19\,km\,s^{-1}$ and surface densities of $\rm \langle \Sigma_{H_2,scouse}\rangle \approx 800\,M_{\odot}\,pc^{-2}$ that are similar across the central molecular gas structure, implying that the observed high dispersion is caused by inter-`cloud' motion between gas peaks. More in-depth analysis in the future has the potential to identify converging flows and relate these to the locations of young clusters.
    
\item The asymmetric gas distribution observed along the bar lanes is also produced in simulations, where it is transient (changing within a dynamical time) and highly time-dependent in nature. Similarly, the resulting distribution of star formation undergoes rapid evolution in the simulations. This is due to a highly variable gas inflow that stems from the clumpiness of the gas distribution that enters the bar and proceeds to the inner structure. This interpretation is consistent with the 7.7$\mu$m PAH distribution that shows a bright neutral gas disk with a brighter ring super-imposed - similar to the time-averaged gas distributions inferred from simulations. 

\item 
There are differences in the overall gas and star formation properties along the gas lanes that are consistent with time evolution of star formation along the ring. However, the onset of star formation along the gas lanes differs significantly along the two lanes. Based on insights from simulations, this could be linked to asymmetric gas inflow or may emerge from a preference for stars to form downstream from the apocenter of the ring. Also high gas inflow rates could play a role as gas entering the ring is not directly converted into stars upon entry but can accumulate along the ring leading to (some) stochasticity in the star formation.

\item Despite the massive ongoing star formation, the 7.7\,$\mu$m distribution does not exhibit many bubbles (especially when compared to galactic disks). This might point to the fact that stellar feedback does not significantly alter the gas distribution, though in-depth analysis of the molecular gas kinematics will be required to confirm this.

\item Based on our analysis we conclude that the massive star formation ongoing in the central 5\,kpc of NGC\,1365 is driven by gas inflow caused by the large-scale stellar bar and that all star formation related properties can be explained without invoking any impact of the AGN on the gas disk.

\end{itemize}

The massive starburst ring in NGC\,1365 provides an excellent opportunity to further our understanding of molecular gas accumulation, collapse and star formation in CMZs. While several of its features appear similar to other star-forming rings observed or seen in simulations, its properties are extreme (especially when compared to the CMZ in the Milky Way). This analysis demonstrates the power of combining high resolution JWST and ALMA data to gain more insights into the physics controlling the star formation process in CMZs. Studying a sample of nearby CMZs can help to overcome limitations to our understanding imposed by the time-varying nature of these structures.

\section*{acknowledgments}
This work has been carried out as part of the PHANGS collaboration. We would like to thank the referee for constructive feedback that helped improve the paper.
ES, TGW, JN acknowledge funding from the European Research Council (ERC) under the European Union’s Horizon 2020 research and innovation programme (grant agreement No. 694343).
This research was supported by the Excellence Cluster ORIGINS which is funded by the Deutsche Forschungsgemeinschaft (DFG, German Research Foundation) under Germany's Excellence Strategy - EXC-2094-390783311. Some of the simulations in this paper have been carried out on the computing facilities of the Computational Center for Particle and Astrophysics (C2PAP). We are grateful for the support by Alexey Krukau and Margarita Petkova through C2PAP.
JS acknowledges support by the Natural Sciences and Engineering Research Council of Canada (NSERC) through a Canadian Institute for Theoretical Astrophysics (CITA) National Fellowship.
MC gratefully acknowledges funding from the DFG through an Emmy Noether Research Group (grant number CH2137/1-1).
COOL Research DAO is a Decentralized Autonomous Organization supporting research in astrophysics aimed at uncovering our cosmic origins.
JMDK gratefully acknowledges funding from the European Research Council (ERC) under the European Union's Horizon 2020 research and innovation programme via the ERC Starting Grant MUSTANG (grant agreement number 714907).
JK gratefully acknowledge funding from the Deutsche Forschungsgemeinschaft (DFG, German Research Foundation)  through the DFG Sachbeihilfe (grant number KR4801/2-1).
EJW, RSK, SCOG acknowledge funding from the Deutsche Forschungsgemeinschaft (DFG, German Research Foundation) -- Project-ID 138713538 -- SFB 881 (``The Milky Way System'', subprojects A1, B1, B2, B8, P1).
FB would like to acknowledge funding from the European Research Council (ERC) under the European Union’s Horizon 2020 research and innovation programme (grant agreement No.726384/Empire).
RSK and MCS thank for support from the European Research Council via the ERC Synergy Grant ``ECOGAL'' (project ID 855130). RSK further thanks for support from the Heidelberg Cluster of Excellence (EXC 2181 - 390900948) ``STRUCTURES'', funded by the German Excellence Strategy, and from the German Ministry for Economic Affairs and Climate Action in the project ``MAINN'' (funding ID 50OO2206). 
KG is supported by the Australian Research Council through the Discovery Early Career Researcher Award (DECRA) Fellowship DE220100766 funded by the Australian Government. 
KG is supported by the Australian Research Council Centre of Excellence for All Sky Astrophysics in 3 Dimensions (ASTRO~3D), through project number CE170100013. 
MQ acknowledges support from the Spanish grant PID2019-106027GA-C44, funded by MCIN/AEI/10.13039/501100011033.
ER and HH acknowledge the support of the Natural Sciences and Engineering Research Council of Canada (NSERC), funding reference number RGPIN-2022-03499.
AKL gratefully acknowledges support by grants 1653300 and 2205628 from the National Science Foundation, by award JWST-GO-02107.009-A, and by a Humboldt Research Award from the Alexander von Humboldt Foundation.
HAP acknowledges support by the National Science and Technology Council of Taiwan under grant 110-2112-M-032-020-MY3.
CE acknowledges funding from the Deutsche Forschungsgemeinschaft (DFG) Sachbeihilfe, grant number BI1546/3-1.
FR acknowledges support from the Knut and Alice Wallenberg Foundation.
AU acknowledges support from the Spanish grants PGC2018-094671-B-I00, funded by MCIN/AEI/10.13039/501100011033 and by ``ERDF A way of making Europe'', and PID2019-108765GB-I00, funded by MCIN/AEI/10.13039/501100011033. 
JPe acknowledges support by the DAOISM grant ANR-21-CE31-0010 and by the Programme National ``Physique et Chimie du Milieu Interstellaire'' (PCMI) of CNRS/INSU with INC/INP, co-funded by CEA and CNES.
SKS acknowledges financial support from the German Research Foundation (DFG) via Sino-German research grant SCHI 536/11-1.

This paper makes use of the following ALMA data: \linebreak
ADS/JAO.ALMA\#2013.1.01161.S. \linebreak 

This work is based on observations made with the NASA/ESA/CSA JWST. The data were obtained from the Mikulski Archive for Space Telescopes (MAST) at the Space Telescope Science Institute, which is operated by the Association of Universities for Research in Astronomy, Inc., under NASA contract NAS 5-03127. The observations are associated with JWST program 2107. The specific observations analyzed can be accessed via \dataset[10.17909/9bdf-jn24]{http://dx.doi.org/10.17909/9bdf-jn24}.

Based on observations collected at the European Southern Observatory under ESO programmes 1100.B-0651 (PHANGS-MUSE; PI: Schinnerer), as well as 094.B-0321 (MAGNUM; PI: Marconi).

\vspace{5mm}
\facilities{ALMA, JWST(MIRI, NIRCam), VLT(MUSE)}


\software{
\texttt{NumPy} \citep{2020Natur.585..357H},
\texttt{Astropy} \citep{2013A&A...558A..33A,2018AJ....156..123A},
\texttt{CASA} \citep{2007ASPC..376..127M},
\texttt{Matplotlib} \citep{2007CSE.....9...90H},
\texttt{APLpy} \citep{2012ascl.soft08017R},
\texttt{CARTA} \citep{2021zndo...3377984C},
\texttt{SCOUSEPY} \citep{Henshaw2016,Henshaw2019},
\texttt{Source Extractor} \citep{1996A&AS..117..393B},
\texttt{yt} \citep{Turk2011},
\texttt{Ramses} \citep{teyssier2002}.
}

\appendix
\section{{\sc ScousePy} Decomposition}
\label{appendix:scouse}

We apply {\sc ScousePy} \citep{Henshaw2016,Henshaw2019} to the ALMA 0.3\arcsec\ CO(2-1) data cube to decompose the emission lines into individual Gaussian components (Fig.\,\ref{fig:scouse}). For the decomposition, we follow the method applied to the PHANGS-ALMA sample in Henshaw et al. (in prep.). We use the `strict\_mask' produced by the PHANGS-ALMA pipeline to guide our decomposition \citep{Leroy2021-Pipeline} as this mask is tailored to contain high confidence emission regions. {\sc ScousePy} divides the spatially masked cube into a Nyquist-sampled set of sub-regions. We set each sub-region to have a width of 35 pixels (i.e.\ $\sim$1.9\arcsec) and require that at least 30\% of the enclosed pixels are not masked for it to be included in the analysis. This results in 850 sub-regions covering the masked area. From each of these sub-regions a spatially averaged spectra is extracted and decomposed using derivative spectroscopy (Henshaw et al. in prep.). The parametric description of each parent sub-region is then passed as an initial guess for the decomposition of all pixels contained within that sub-region. This process depends on a set of tolerance conditions that control the modelling, either by measuring the properties of individual components (e.g.\ their minimum signal-to-noise, which we set to 3, and width, which we set to 1 channel, respectively) or how they compare to the closest matching component in the parent sub-region\footnote{Specifically, the set of tolerance conditions used in this study are $\mathrm{tol}=[4.0, 3.0, 0.5, 5.0, 5.0, 0.5]$ and we refer the reader to \cite{Henshaw2016} for further details.}.

To improve the decomposition we impose the quality control procedure outlined in Henshaw et al. (in prep.). Briefly, the quality control process involves two main steps: i) flagging problematic spectra; ii) seeking alternative solutions to problematic spectra. Spectra with no associated model solution (typically those which violate the tolerance conditions outlined above), those where the parameter uncertainties are high post-decomposition, or those for which the model solutions differ substantially from that of their surrounding neighbours are flagged. In total, $\sim28\%$ of the spectra in the masked region were flagged. The vast majority of these flags ($\sim86\%$) are triggered by spectra with no associated model solution. Broadly speaking, these spectra are located at the edge of the mapped region, where the signal-to-noise ratio is low. The next step is to seek alternative solutions for the flagged spectra. We do this in two ways. First, we attempt to take advantage of the Nyquist sampling of sub-regions, which can lead to alternative models being available for problematic spectra. Second, where alternatives are not available, {\sc ScousePy} performs a neighbour-based refit of the spectrum (where we relaxed the condition on the amplitude to also include components of marginal significance, i.e.\ those with signal-to-noise $>2$). After quality control, $<7\%$ of the spectra remain flagged, where again most of which come from pixels where {\sc ScousePy} was unable to find a suitable solution (top-left panel Fig.\,\ref{fig:scouse}). 


\begin{thebibliography}{}
\expandafter\ifx\csname natexlab\endcsname\relax\def\natexlab#1{#1}\fi
\providecommand{\url}[1]{\href{#1}{#1}}
\providecommand{\dodoi}[1]{doi:~\href{http://doi.org/#1}{\nolinkurl{#1}}}
\providecommand{\doeprint}[1]{\href{http://ascl.net/#1}{\nolinkurl{http://ascl.net/#1}}}
\providecommand{\doarXiv}[1]{\href{https://arxiv.org/abs/#1}{\nolinkurl{https://arxiv.org/abs/#1}}}

\bibitem[{{Agertz} \& {Kravtsov}(2015)}]{agertz+2015}
{Agertz}, O., \& {Kravtsov}, A.~V. 2015, \apj, 804, 18,
  \dodoi{10.1088/0004-637X/804/1/18}

\bibitem[{{Agertz} {et~al.}(2021){Agertz}, {Renaud}, {Feltzing}, {Read},
  {Ryde}, {Andersson}, {Rey}, {Bensby}, \& {Feuillet}}]{agertz+2021}
{Agertz}, O., {Renaud}, F., {Feltzing}, S., {et~al.} 2021, \mnras, 503, 5826,
  \dodoi{10.1093/mnras/stab322}

\bibitem[{{Alonso-Herrero} {et~al.}(2012){Alonso-Herrero},
  {S{\'a}nchez-Portal}, {Ramos Almeida}, {Pereira-Santaella}, {Esquej},
  {Garc{\'\i}a-Burillo}, {Castillo}, {Gonz{\'a}lez-Mart{\'\i}n}, {Levenson},
  {Hatziminaoglou}, {Acosta-Pulido}, {Gonz{\'a}lez-Serrano}, {Povi{\'c}},
  {Packham}, \& {P{\'e}rez-Garc{\'\i}a}}]{Alonso-Herrero2012}
{Alonso-Herrero}, A., {S{\'a}nchez-Portal}, M., {Ramos Almeida}, C., {et~al.}
  2012, \mnras, 425, 311, \dodoi{10.1111/j.1365-2966.2012.21464.x}

\bibitem[{{Alonso-Herrero} {et~al.}(2020){Alonso-Herrero}, {Pereira-Santaella},
  {Rigopoulou}, {Garc{\'\i}a-Bernete}, {Garc{\'\i}a-Burillo},
  {Dom{\'\i}nguez-Fern{\'a}ndez}, {Combes}, {Davies}, {D{\'\i}az-Santos},
  {Esparza-Arredondo}, {Gonz{\'a}lez-Mart{\'\i}n}, {Hern{\'a}n-Caballero},
  {Hicks}, {H{\"o}nig}, {Levenson}, {Ramos Almeida}, {Roche}, \&
  {Rosario}}]{alonso-herrero2020}
{Alonso-Herrero}, A., {Pereira-Santaella}, M., {Rigopoulou}, D., {et~al.} 2020,
  \aap, 639, A43, \dodoi{10.1051/0004-6361/202037642}

\bibitem[{{Anand} {et~al.}(2021{\natexlab{a}}){Anand}, {Rizzi}, {Tully},
  {Shaya}, {Karachentsev}, {Makarov}, {Makarova}, {Wu}, {Dolphin}, \&
  {Kourkchi}}]{Anand2021a}
{Anand}, G.~S., {Rizzi}, L., {Tully}, R.~B., {et~al.} 2021{\natexlab{a}}, \aj,
  162, 80, \dodoi{10.3847/1538-3881/ac0440}

\bibitem[{{Anand} {et~al.}(2021{\natexlab{b}}){Anand}, {Lee}, {Van Dyk},
  {Leroy}, {Rosolowsky}, {Schinnerer}, {Larson}, {Kourkchi}, {Kreckel},
  {Scheuermann}, {Rizzi}, {Thilker}, {Tully}, {Bigiel}, {Blanc}, {Boquien},
  {Chandar}, {Dale}, {Emsellem}, {Deger}, {Glover}, {Grasha}, {Groves}, {S.
  Klessen}, {Kruijssen}, {Querejeta}, {S{\'a}nchez-Bl{\'a}zquez}, {Schruba},
  {Turner}, {Ubeda}, {Williams}, \& {Whitmore}}]{Anand2021b}
{Anand}, G.~S., {Lee}, J.~C., {Van Dyk}, S.~D., {et~al.} 2021{\natexlab{b}},
  \mnras, 501, 3621, \dodoi{10.1093/mnras/staa3668}

\bibitem[{{Armillotta} {et~al.}(2019){Armillotta}, {Krumholz}, {Di Teodoro}, \&
  {McClure-Griffiths}}]{Armillotta2019}
{Armillotta}, L., {Krumholz}, M.~R., {Di Teodoro}, E.~M., \&
  {McClure-Griffiths}, N.~M. 2019, \mnras, 490, 4401,
  \dodoi{10.1093/mnras/stz2880}

\bibitem[{{Astropy Collaboration} {et~al.}(2013){Astropy Collaboration},
  {Robitaille}, {Tollerud}, {Greenfield}, {Droettboom}, {Bray}, {Aldcroft},
  {Davis}, {Ginsburg}, {Price-Whelan}, {Kerzendorf}, {Conley}, {Crighton},
  {Barbary}, {Muna}, {Ferguson}, {Grollier}, {Parikh}, {Nair}, {Unther},
  {Deil}, {Woillez}, {Conseil}, {Kramer}, {Turner}, {Singer}, {Fox}, {Weaver},
  {Zabalza}, {Edwards}, {Azalee Bostroem}, {Burke}, {Casey}, {Crawford},
  {Dencheva}, {Ely}, {Jenness}, {Labrie}, {Lim}, {Pierfederici}, {Pontzen},
  {Ptak}, {Refsdal}, {Servillat}, \& {Streicher}}]{2013A&A...558A..33A}
{Astropy Collaboration}, {Robitaille}, T.~P., {Tollerud}, E.~J., {et~al.} 2013,
  \aap, 558, A33, \dodoi{10.1051/0004-6361/201322068}

\bibitem[{{Astropy Collaboration} {et~al.}(2018){Astropy Collaboration},
  {Price-Whelan}, {Sip{\H{o}}cz}, {G{\"u}nther}, {Lim}, {Crawford}, {Conseil},
  {Shupe}, {Craig}, {Dencheva}, {Ginsburg}, {VanderPlas}, {Bradley},
  {P{\'e}rez-Su{\'a}rez}, {de Val-Borro}, {Aldcroft}, {Cruz}, {Robitaille},
  {Tollerud}, {Ardelean}, {Babej}, {Bach}, {Bachetti}, {Bakanov}, {Bamford},
  {Barentsen}, {Barmby}, {Baumbach}, {Berry}, {Biscani}, {Boquien}, {Bostroem},
  {Bouma}, {Brammer}, {Bray}, {Breytenbach}, {Buddelmeijer}, {Burke},
  {Calderone}, {Cano Rodr{\'\i}guez}, {Cara}, {Cardoso}, {Cheedella}, {Copin},
  {Corrales}, {Crichton}, {D'Avella}, {Deil}, {Depagne}, {Dietrich}, {Donath},
  {Droettboom}, {Earl}, {Erben}, {Fabbro}, {Ferreira}, {Finethy}, {Fox},
  {Garrison}, {Gibbons}, {Goldstein}, {Gommers}, {Greco}, {Greenfield},
  {Groener}, {Grollier}, {Hagen}, {Hirst}, {Homeier}, {Horton}, {Hosseinzadeh},
  {Hu}, {Hunkeler}, {Ivezi{\'c}}, {Jain}, {Jenness}, {Kanarek}, {Kendrew},
  {Kern}, {Kerzendorf}, {Khvalko}, {King}, {Kirkby}, {Kulkarni}, {Kumar},
  {Lee}, {Lenz}, {Littlefair}, {Ma}, {Macleod}, {Mastropietro}, {McCully},
  {Montagnac}, {Morris}, {Mueller}, {Mumford}, {Muna}, {Murphy}, {Nelson},
  {Nguyen}, {Ninan}, {N{\"o}the}, {Ogaz}, {Oh}, {Parejko}, {Parley}, {Pascual},
  {Patil}, {Patil}, {Plunkett}, {Prochaska}, {Rastogi}, {Reddy Janga},
  {Sabater}, {Sakurikar}, {Seifert}, {Sherbert}, {Sherwood-Taylor}, {Shih},
  {Sick}, {Silbiger}, {Singanamalla}, {Singer}, {Sladen}, {Sooley},
  {Sornarajah}, {Streicher}, {Teuben}, {Thomas}, {Tremblay}, {Turner},
  {Terr{\'o}n}, {van Kerkwijk}, {de la Vega}, {Watkins}, {Weaver}, {Whitmore},
  {Woillez}, {Zabalza}, \& {Astropy Contributors}}]{2018AJ....156..123A}
{Astropy Collaboration}, {Price-Whelan}, A.~M., {Sip{\H{o}}cz}, B.~M., {et~al.}
  2018, \aj, 156, 123, \dodoi{10.3847/1538-3881/aabc4f}

\bibitem[{{Athanassoula}(1992)}]{Athanassoula1992}
{Athanassoula}, E. 1992, \mnras, 259, 345, \dodoi{10.1093/mnras/259.2.345}

\bibitem[{{Behrens} {et~al.}(2022){Behrens}, {Mangum}, {Holdship}, {Viti},
  {Harada}, {Martin}, {Sakamoto}, {Muller}, {Tanaka}, {Nakanishi},
  {Herrero-Illana}, {Yoshimura}, {Aladro}, {Colzi}, {Emig}, {Henkel}, {Huang},
  {Humire}, {Meier}, \& {Rivilla}}]{Behrens2022}
{Behrens}, E., {Mangum}, J.~G., {Holdship}, J., {et~al.} 2022, arXiv e-prints,
  arXiv:2209.06244.
\newblock \doarXiv{2209.06244}

\bibitem[{{Belfiore} {et~al.}(2022){Belfiore}, {Santoro}, {Groves},
  {Schinnerer}, {Kreckel}, {Glover}, {Klessen}, {Emsellem}, {Blanc}, {Congiu},
  {Barnes}, {Boquien}, {Chevance}, {Dale}, {Kruijssen}, {Leroy}, {Pan},
  {Pessa}, {Schruba}, \& {Williams}}]{Belfiore2022}
{Belfiore}, F., {Santoro}, F., {Groves}, B., {et~al.} 2022, \aap, 659, A26,
  \dodoi{10.1051/0004-6361/202141859}

\bibitem[{{Bertin} \& {Arnouts}(1996)}]{1996A&AS..117..393B}
{Bertin}, E., \& {Arnouts}, S. 1996, \aaps, 117, 393,
  \dodoi{10.1051/aas:1996164}

\bibitem[{{Bertoldi} \& {McKee}(1992)}]{bertoldi1992}
{Bertoldi}, F., \& {McKee}, C.~F. 1992, \apj, 395, 140, \dodoi{10.1086/171638}

\bibitem[{{B{\"o}ker} {et~al.}(2008){B{\"o}ker}, {Falc{\'o}n-Barroso},
  {Schinnerer}, {Knapen}, \& {Ryder}}]{boeker2008}
{B{\"o}ker}, T., {Falc{\'o}n-Barroso}, J., {Schinnerer}, E., {Knapen}, J.~H.,
  \& {Ryder}, S. 2008, \aj, 135, 479, \dodoi{10.1088/0004-6256/135/2/479}

\bibitem[{{Bolatto} {et~al.}(2013){Bolatto}, {Wolfire}, \&
  {Leroy}}]{bolatto2013}
{Bolatto}, A.~D., {Wolfire}, M., \& {Leroy}, A.~K. 2013, \araa, 51, 207,
  \dodoi{10.1146/annurev-astro-082812-140944}

\bibitem[{{Buta} {et~al.}(2015){Buta}, {Sheth}, {Athanassoula}, {Bosma},
  {Knapen}, {Laurikainen}, {Salo}, {Elmegreen}, {Ho}, {Zaritsky}, {Courtois},
  {Hinz}, {Mu{\~n}oz-Mateos}, {Kim}, {Regan}, {Gadotti}, {Gil de Paz}, {Laine},
  {Men{\'e}ndez-Delmestre}, {Comer{\'o}n}, {Erroz Ferrer}, {Seibert},
  {Mizusawa}, {Holwerda}, \& {Madore}}]{Buta2015}
{Buta}, R.~J., {Sheth}, K., {Athanassoula}, E., {et~al.} 2015, \apjs, 217, 32,
  \dodoi{10.1088/0067-0049/217/2/32}

\bibitem[{{Callanan} {et~al.}(2021){Callanan}, {Longmore}, {Kruijssen},
  {Schruba}, {Ginsburg}, {Krumholz}, {Bastian}, {Alves}, {Henshaw}, {Knapen},
  \& {Chevance}}]{Callanan2021}
{Callanan}, D., {Longmore}, S.~N., {Kruijssen}, J.~M.~D., {et~al.} 2021,
  \mnras, 505, 4310, \dodoi{10.1093/mnras/stab1527}

\bibitem[{{Chevance} {et~al.}(2020){Chevance}, {Kruijssen}, {Hygate},
  {Schruba}, {Longmore}, {Groves}, {Henshaw}, {Herrera}, {Hughes}, {Jeffreson},
  {Lang}, {Leroy}, {Meidt}, {Pety}, {Razza}, {Rosolowsky}, {Schinnerer},
  {Bigiel}, {Blanc}, {Emsellem}, {Faesi}, {Glover}, {Haydon}, {Ho}, {Kreckel},
  {Lee}, {Liu}, {Querejeta}, {Saito}, {Sun}, {Usero}, \&
  {Utomo}}]{chevance2020}
{Chevance}, M., {Kruijssen}, J.~M.~D., {Hygate}, A. P.~S., {et~al.} 2020,
  \mnras, 493, 2872, \dodoi{10.1093/mnras/stz3525}

\bibitem[{{Combes} {et~al.}(2019){Combes}, {Garc{\'\i}a-Burillo}, {Audibert},
  {Hunt}, {Eckart}, {Aalto}, {Casasola}, {Boone}, {Krips}, {Viti}, {Sakamoto},
  {Muller}, {Dasyra}, {van der Werf}, \& {Martin}}]{combes2019}
{Combes}, F., {Garc{\'\i}a-Burillo}, S., {Audibert}, A., {et~al.} 2019, \aap,
  623, A79, \dodoi{10.1051/0004-6361/201834560}

\bibitem[{{Comer{\'o}n} {et~al.}(2014){Comer{\'o}n}, {Salo}, {Laurikainen},
  {Knapen}, {Buta}, {Herrera-Endoqui}, {Laine}, {Holwerda}, {Sheth}, {Regan},
  {Hinz}, {Mu{\~n}oz-Mateos}, {Gil de Paz}, {Men{\'e}ndez-Delmestre},
  {Seibert}, {Mizusawa}, {Kim}, {Erroz-Ferrer}, {Gadotti}, {Athanassoula},
  {Bosma}, \& {Ho}}]{comeron2014}
{Comer{\'o}n}, S., {Salo}, H., {Laurikainen}, E., {et~al.} 2014, \aap, 562,
  A121, \dodoi{10.1051/0004-6361/201321633}

\bibitem[{{Comrie} {et~al.}(2021){Comrie}, {Wang}, {Hsu}, {Moraghan}, {Harris},
  {Pang}, {Pi{\'n}ska}, {Chiang}, {Chang}, {Hwang}, {Jan}, {Lin}, \&
  {Simmonds}}]{2021zndo...3377984C}
{Comrie}, A., {Wang}, K.-S., {Hsu}, S.-C., {et~al.} 2021, {CARTA: The Cube
  Analysis and Rendering Tool for Astronomy}, 2.0.0, Zenodo,  Zenodo,
  \dodoi{10.5281/zenodo.3377984}

\bibitem[{{Davis} {et~al.}(2022){Davis}, {Gensior}, {Bureau}, {Cappellari},
  {Choi}, {Elford}, {Kruijssen}, {Lelli}, {Liang}, {Liu}, {Ruffa}, {Saito},
  {Sarzi}, {Schruba}, \& {Williams}}]{davis2022}
{Davis}, T.~A., {Gensior}, J., {Bureau}, M., {et~al.} 2022, \mnras, 512, 1522,
  \dodoi{10.1093/mnras/stac600}

\bibitem[{{Edmunds} {et~al.}(1988){Edmunds}, {Taylor}, \&
  {Turtle}}]{edmunds1988}
{Edmunds}, M.~G., {Taylor}, K., \& {Turtle}, A.~J. 1988, \mnras, 234, 155,
  \dodoi{10.1093/mnras/234.1.155}

\bibitem[{{Egusa} {et~al.}(2022){Egusa}, {Gao}, {Morokuma-Matsui}, {Liu}, \&
  {Maeda}}]{Egusa2022}
{Egusa}, F., {Gao}, Y., {Morokuma-Matsui}, K., {Liu}, G., \& {Maeda}, F. 2022,
  \apj, 935, 64, \dodoi{10.3847/1538-4357/ac8050}

\bibitem[{{Eibensteiner} {et~al.}(2022){Eibensteiner}, {Barnes}, {Bigiel},
  {Schinnerer}, {Liu}, {Meier}, {Usero}, {Leroy}, {Rosolowsky}, {Puschnig},
  {Lazar}, {Pety}, {Lopez}, {Emsellem}, {Be{\v{s}}li{\'c}}, {Querejeta},
  {Murphy}, {den Brok}, {Schruba}, {Chevance}, {Glover}, {Gao}, {Grasha},
  {Hassani}, {Henshaw}, {Jimenez-Donaire}, {Klessen}, {Kruijssen}, {Pan},
  {Saito}, {Sormani}, {Teng}, \& {Williams}}]{Eibensteiner2022}
{Eibensteiner}, C., {Barnes}, A.~T., {Bigiel}, F., {et~al.} 2022, \aap, 659,
  A173, \dodoi{10.1051/0004-6361/202142624}

\bibitem[{{Elmegreen}(1994)}]{Elmegreen1994}
{Elmegreen}, B.~G. 1994, \apjl, 425, L73, \dodoi{10.1086/187313}

\bibitem[{{Elmegreen} {et~al.}(2009){Elmegreen}, {Galliano}, \&
  {Alloin}}]{Elmegreen2009}
{Elmegreen}, B.~G., {Galliano}, E., \& {Alloin}, D. 2009, \apj, 703, 1297,
  \dodoi{10.1088/0004-637X/703/2/1297}

\bibitem[{{Emsellem} {et~al.}(1994){Emsellem}, {Monnet}, \&
  {Bacon}}]{emsellem+1994}
{Emsellem}, E., {Monnet}, G., \& {Bacon}, R. 1994, \aap, 285, 723

\bibitem[{{Emsellem} {et~al.}(2015{\natexlab{a}}){Emsellem}, {Renaud},
  {Bournaud}, {Elmegreen}, {Combes}, \& {Gabor}}]{Emsellem2015}
{Emsellem}, E., {Renaud}, F., {Bournaud}, F., {et~al.} 2015{\natexlab{a}},
  \mnras, 446, 2468, \dodoi{10.1093/mnras/stu2209}

\bibitem[{{Emsellem} {et~al.}(2015{\natexlab{b}}){Emsellem}, {Renaud},
  {Bournaud}, {Elmegreen}, {Combes}, \& {Gabor}}]{emsellem+2015}
---. 2015{\natexlab{b}}, \mnras, 446, 2468, \dodoi{10.1093/mnras/stu2209}

\bibitem[{{Emsellem} {et~al.}(2022){Emsellem}, {Schinnerer}, {Santoro},
  {Belfiore}, {Pessa}, {McElroy}, {Blanc}, {Congiu}, {Groves}, {Ho}, {Kreckel},
  {Razza}, {Sanchez-Blazquez}, {Egorov}, {Faesi}, {Klessen}, {Leroy}, {Meidt},
  {Querejeta}, {Rosolowsky}, {Scheuermann}, {Anand}, {Barnes},
  {Be{\v{s}}li{\'c}}, {Bigiel}, {Boquien}, {Cao}, {Chevance}, {Dale},
  {Eibensteiner}, {Glover}, {Grasha}, {Henshaw}, {Hughes}, {Koch}, {Kruijssen},
  {Lee}, {Liu}, {Pan}, {Pety}, {Saito}, {Sandstrom}, {Schruba}, {Sun},
  {Thilker}, {Usero}, {Watkins}, \& {Williams}}]{Emsellem2022}
{Emsellem}, E., {Schinnerer}, E., {Santoro}, F., {et~al.} 2022, \aap, 659,
  A191, \dodoi{10.1051/0004-6361/202141727}

\bibitem[{{Fazeli} {et~al.}(2019){Fazeli}, {Busch}, {Valencia-S.}, {Eckart},
  {Zaja{\v{c}}ek}, {Combes}, \& {Garc{\'\i}a-Burillo}}]{Fazeli2019}
{Fazeli}, N., {Busch}, G., {Valencia-S.}, M., {et~al.} 2019, \aap, 622, A128,
  \dodoi{10.1051/0004-6361/201834255}

\bibitem[{{Forbes} \& {Norris}(1998)}]{Forbes1998}
{Forbes}, D.~A., \& {Norris}, R.~P. 1998, \mnras, 300, 757,
  \dodoi{10.1046/j.1365-8711.1998.01940.x}

\bibitem[{{Galliano} {et~al.}(2008){Galliano}, {Alloin}, {Pantin}, {Granato},
  {Delva}, {Silva}, {Lagage}, \& {Panuzzo}}]{galliano2008}
{Galliano}, E., {Alloin}, D., {Pantin}, E., {et~al.} 2008, \aap, 492, 3,
  \dodoi{10.1051/0004-6361:20077621}

\bibitem[{{Galliano} {et~al.}(2012){Galliano}, {Kissler-Patig}, {Alloin}, \&
  {Telles}}]{galliano2012}
{Galliano}, E., {Kissler-Patig}, M., {Alloin}, D., \& {Telles}, E. 2012, \aap,
  545, A10, \dodoi{10.1051/0004-6361/201218812}

\bibitem[{{Galliano} {et~al.}(2005){Galliano}, {Pantin}, {Alloin}, \&
  {Lagage}}]{Galliano2005}
{Galliano}, E., {Pantin}, E., {Alloin}, D., \& {Lagage}, P.~O. 2005, \mnras,
  363, L1, \dodoi{10.1111/j.1745-3933.2005.00064.x}

\bibitem[{{Galliano} {et~al.}(2018){Galliano}, {Galametz}, \&
  {Jones}}]{galliano2018}
{Galliano}, F., {Galametz}, M., \& {Jones}, A.~P. 2018, \araa, 56, 673,
  \dodoi{10.1146/annurev-astro-081817-051900}

\bibitem[{{Gao} {et~al.}(2021){Gao}, {Egusa}, {Liu}, {Kohno}, {Bao},
  {Morokuma-Matsui}, {Kong}, \& {Chen}}]{Gao2021}
{Gao}, Y., {Egusa}, F., {Liu}, G., {et~al.} 2021, \apj, 913, 139,
  \dodoi{10.3847/1538-4357/abf738}

\bibitem[{{Gensior} {et~al.}(2020){Gensior}, {Kruijssen}, \&
  {Keller}}]{gensior2020}
{Gensior}, J., {Kruijssen}, J.~M.~D., \& {Keller}, B.~W. 2020, \mnras, 495,
  199, \dodoi{10.1093/mnras/staa1184}

\bibitem[{{Girichidis} {et~al.}(2020){Girichidis}, {Offner}, {Kritsuk},
  {Klessen}, {Hennebelle}, {Kruijssen}, {Krause}, {Glover}, \&
  {Padovani}}]{Girichidis2020}
{Girichidis}, P., {Offner}, S. S.~R., {Kritsuk}, A.~G., {et~al.} 2020, \ssr,
  216, 68, \dodoi{10.1007/s11214-020-00693-8}

\bibitem[{{Habibi} {et~al.}(2014){Habibi}, {Stolte}, \& {Harfst}}]{habibi2014}
{Habibi}, M., {Stolte}, A., \& {Harfst}, S. 2014, \aap, 566, A6,
  \dodoi{10.1051/0004-6361/201323030}

\bibitem[{{Harris} {et~al.}(2020){Harris}, {Millman}, {van der Walt},
  {Gommers}, {Virtanen}, {Cour\ napeau}, {Wieser}, {Taylor}, {Berg}, {Smith},
  {Kern}, {Picus}, {van Kerkwijk}, {Brett}, {Haldane}, {del R{\'\i}o}, {Wiebe\
  }, {Peterson}, {G{\'e}rard-Marchant}, {Sheppard}, {Reddy}, {Weckesser},
  {Abbasi}, {Gohlke}, \& {Oliphant}}]{2020Natur.585..357H}
{Harris}, C.~R., {Millman}, K.~J., {van der Walt}, S.~J., {et~al.} 2020, \nat,
  585, 357, \dodoi{10.1038/s41586-020-2649-2}

\bibitem[{{Hassani} {et~al.}(2022){Hassani}, {Rosolowsky}, {Leroy}, {Boquien},
  {Lee}, {Barnes}, {Belfiore}, {Bigiel}, {Cao}, {Chevance}, {Dale}, {Egorov},
  {Emsellem}, {Faesi}, {Grasha}, {Kim}, {Klessen}, {Kreckel}, {Kruijssen},
  {Larson}, {Meidt}, {Sandstrom}, {Schinnerer}, {Thilker}, {Watkins},
  {Whitmore}, \& {Williams}}]{HASSANI_PHANGSJWST}
{Hassani}, H., {Rosolowsky}, E., {Leroy}, A.~K., {et~al.} 2022, arXiv e-prints,
  arXiv:2212.01526.
\newblock \doarXiv{2212.01526}

\bibitem[{{Hatchfield} {et~al.}(2021){Hatchfield}, {Sormani}, {Tress},
  {Battersby}, {Smith}, {Glover}, \& {Klessen}}]{hatchfield2021}
{Hatchfield}, H.~P., {Sormani}, M.~C., {Tress}, R.~G., {et~al.} 2021, \apj,
  922, 79, \dodoi{10.3847/1538-4357/ac1e89}

\bibitem[{{Henshaw} {et~al.}(2022){Henshaw}, {Barnes}, {Battersby}, {Ginsburg},
  {Sormani}, \& {Walker}}]{henshaw2022}
{Henshaw}, J.~D., {Barnes}, A.~T., {Battersby}, C., {et~al.} 2022, arXiv
  e-prints, arXiv:2203.11223.
\newblock \doarXiv{2203.11223}

\bibitem[{{Henshaw} {et~al.}(2016){Henshaw}, {Longmore}, {Kruijssen}, {Davies},
  {Bally}, {Barnes}, {Battersby}, {Burton}, {Cunningham}, {Dale}, {Ginsburg},
  {Immer}, {Jones}, {Kendrew}, {Mills}, {Molinari}, {Moore}, {Ott}, {Pillai},
  {Rathborne}, {Schilke}, {Schmiedeke}, {Testi}, {Walker}, {Walsh}, \&
  {Zhang}}]{Henshaw2016}
{Henshaw}, J.~D., {Longmore}, S.~N., {Kruijssen}, J.~M.~D., {et~al.} 2016,
  \mnras, 457, 2675, \dodoi{10.1093/mnras/stw121}

\bibitem[{{Henshaw} {et~al.}(2019){Henshaw}, {Ginsburg}, {Haworth}, {Longmore},
  {Kruijssen}, {Mills}, {Sokolov}, {Walker}, {Barnes}, {Contreras}, {Bally},
  {Battersby}, {Beuther}, {Butterfield}, {Dale}, {Henning}, {Jackson},
  {Kauffmann}, {Pillai}, {Ragan}, {Riener}, \& {Zhang}}]{Henshaw2019}
{Henshaw}, J.~D., {Ginsburg}, A., {Haworth}, T.~J., {et~al.} 2019, \mnras, 485,
  2457, \dodoi{10.1093/mnras/stz471}

\bibitem[{{Henshaw} {et~al.}(2020){Henshaw}, {Kruijssen}, {Longmore}, {Riener},
  {Leroy}, {Rosolowsky}, {Ginsburg}, {Battersby}, {Chevance}, {Meidt},
  {Glover}, {Hughes}, {Kainulainen}, {Klessen}, {Schinnerer}, {Schruba},
  {Beuther}, {Bigiel}, {Blanc}, {Emsellem}, {Henning}, {Herrera}, {Koch},
  {Pety}, {Ragan}, \& {Sun}}]{Henshaw2020}
{Henshaw}, J.~D., {Kruijssen}, J.~M.~D., {Longmore}, S.~N., {et~al.} 2020,
  Nature Astronomy, 4, 1064, \dodoi{10.1038/s41550-020-1126-z}

\bibitem[{{Herrera-Endoqui} {et~al.}(2015){Herrera-Endoqui},
  {D{\'\i}az-Garc{\'\i}a}, {Laurikainen}, \& {Salo}}]{herrera-endoqui2015}
{Herrera-Endoqui}, M., {D{\'\i}az-Garc{\'\i}a}, S., {Laurikainen}, E., \&
  {Salo}, H. 2015, \aap, 582, A86, \dodoi{10.1051/0004-6361/201526047}

\bibitem[{{Hjelm} \& {Lindblad}(1996)}]{hjelm1996}
{Hjelm}, M., \& {Lindblad}, P.~O. 1996, \aap, 305, 727

\bibitem[{{Hunter}(2007)}]{2007CSE.....9...90H}
{Hunter}, J.~D. 2007, Computing in Science and Engineering, 9, 90,
  \dodoi{10.1109/MCSE.2007.55}

\bibitem[{{Jones} \& {Jones}(1980)}]{Jones1980}
{Jones}, J.~E., \& {Jones}, B.~J.~T. 1980, \mnras, 191, 685,
  \dodoi{10.1093/mnras/191.4.685}

\bibitem[{{Jorsater} {et~al.}(1984){Jorsater}, {Lindblad}, \&
  {Boksenberg}}]{jorsater1984}
{Jorsater}, S., {Lindblad}, P.~O., \& {Boksenberg}, A. 1984, \aap, 140, 288

\bibitem[{{Kim} {et~al.}(2021){Kim}, {Chevance}, {Kruijssen}, {Schruba},
  {Sandstrom}, {Barnes}, {Bigiel}, {Blanc}, {Cao}, {Dale}, {Faesi}, {Glover},
  {Grasha}, {Groves}, {Herrera}, {Klessen}, {Kreckel}, {Lee}, {Leroy}, {Pety},
  {Querejeta}, {Schinnerer}, {Sun}, {Usero}, {Ward}, \& {Williams}}]{kim2021}
{Kim}, J., {Chevance}, M., {Kruijssen}, J.~M.~D., {et~al.} 2021, \mnras, 504,
  487, \dodoi{10.1093/mnras/stab878}

\bibitem[{{Kim} {et~al.}(2022{\natexlab{a}}){Kim}, {Chevance}, {Kruijssen},
  {Leroy}, {Schruba}, {Barnes}, {Bigiel}, {Blanc}, {Cao}, {Congiu}, {Dale},
  {Faesi}, {Glover}, {Grasha}, {Groves}, {Hughes}, {Klessen}, {Kreckel},
  {McElroy}, {Pan}, {Pety}, {Querejeta}, {Razza}, {Rosolowsky}, {Saito},
  {Schinnerer}, {Sun}, {Tomi{\v{c}}i{\'c}}, {Usero}, \& {Williams}}]{kim2022}
---. 2022{\natexlab{a}}, \mnras, 516, 3006, \dodoi{10.1093/mnras/stac2339}

\bibitem[{{Kim} {et~al.}(2022{\natexlab{b}}){Kim}, {Chevance}, {Kruijssen},
  {Barnes}, {Bigiel}, {Blanc}, {Boquien}, {Cao}, {Congiu}, {Dale}, {Egorov},
  {Faesi}, {Glover}, {Grasha}, {Groves}, {Hassani}, {Hughes}, {Klessen},
  {Kreckel}, {Larson}, {Lee}, {Leroy}, {Liu}, {Longmore}, {Meidt}, {Pan},
  {Pety}, {Querejeta}, {Rosolowsky}, {Saito}, {Sandstrom}, {Schinnerer},
  {Smith}, {Usero}, {Watkins}, \& {Williams}}]{KIM_PHANGSJWST}
---. 2022{\natexlab{b}}, arXiv e-prints, arXiv:2211.15698.
\newblock \doarXiv{2211.15698}

\bibitem[{{Knapen} {et~al.}(2000){Knapen}, {Shlosman}, \&
  {Peletier}}]{Knapen2000}
{Knapen}, J.~H., {Shlosman}, I., \& {Peletier}, R.~F. 2000, \apj, 529, 93,
  \dodoi{10.1086/308266}

\bibitem[{{Kormendy} \& {Kennicutt}(2004)}]{Kormendy2004}
{Kormendy}, J., \& {Kennicutt}, Robert~C., J. 2004, \araa, 42, 603,
  \dodoi{10.1146/annurev.astro.42.053102.134024}

\bibitem[{{Kristen} {et~al.}(1997){Kristen}, {Jorsater}, {Lindblad}, \&
  {Boksenberg}}]{Kristen1997}
{Kristen}, H., {Jorsater}, S., {Lindblad}, P.~O., \& {Boksenberg}, A. 1997,
  \aap, 328, 483

\bibitem[{{Krugel} \& {Tutukov}(1993)}]{Krugel1993}
{Krugel}, E., \& {Tutukov}, A.~V. 1993, \aap, 275, 416

\bibitem[{{Kruijssen} {et~al.}(2014){Kruijssen}, {Longmore}, {Elmegreen},
  {Murray}, {Bally}, {Testi}, \& {Kennicutt}}]{Kruijssen2014}
{Kruijssen}, J.~M.~D., {Longmore}, S.~N., {Elmegreen}, B.~G., {et~al.} 2014,
  \mnras, 440, 3370, \dodoi{10.1093/mnras/stu494}

\bibitem[{{Kruijssen} {et~al.}(2019){Kruijssen}, {Schruba}, {Chevance},
  {Longmore}, {Hygate}, {Haydon}, {McLeod}, {Dalcanton}, {Tacconi}, \& {van
  Dishoeck}}]{kruijssen2019}
{Kruijssen}, J.~M.~D., {Schruba}, A., {Chevance}, M., {et~al.} 2019, \nat, 569,
  519, \dodoi{10.1038/s41586-019-1194-3}

\bibitem[{{Krumholz} {et~al.}(2017){Krumholz}, {Kruijssen}, \&
  {Crocker}}]{krumholz2017}
{Krumholz}, M.~R., {Kruijssen}, J.~M.~D., \& {Crocker}, R.~M. 2017, \mnras,
  466, 1213, \dodoi{10.1093/mnras/stw3195}

\bibitem[{{Krumholz} {et~al.}(2009){Krumholz}, {McKee}, \&
  {Tumlinson}}]{krumholz+2009}
{Krumholz}, M.~R., {McKee}, C.~F., \& {Tumlinson}, J. 2009, \apj, 699, 850,
  \dodoi{10.1088/0004-637X/699/1/850}

\bibitem[{{Lang} {et~al.}(2020){Lang}, {Meidt}, {Rosolowsky}, {Nofech},
  {Schinnerer}, {Leroy}, {Emsellem}, {Pessa}, {Glover}, {Groves}, {Hughes},
  {Kruijssen}, {Querejeta}, {Schruba}, {Bigiel}, {Blanc}, {Chevance},
  {Colombo}, {Faesi}, {Henshaw}, {Herrera}, {Liu}, {Pety}, {Puschnig}, {Saito},
  {Sun}, \& {Usero}}]{lang2020}
{Lang}, P., {Meidt}, S.~E., {Rosolowsky}, E., {et~al.} 2020, \apj, 897, 122,
  \dodoi{10.3847/1538-4357/ab9953}

\bibitem[{{Lee} {et~al.}(2022{\natexlab{a}}){Lee}, {Sandstrom}, {Leroy},
  {Thilker}, {Schinnerer}, {Rosolowsky}, {Larson}, {Egorov}, {Williams},
  {Schmidt}, {Emsellem}, {Anand}, {Barnes}, {Belfiore}, {Beslic}, {Bigiel},
  {Blanc}, {Bolatto}, {Boquien}, {den Brok}, {Cao}, {Chandar}, {Chastenet},
  {Chevance}, {Chiang}, {Congiu}, {Dale}, {Deger}, {Eibensteiner}, {Faesi},
  {Glover}, {Grasha}, {Groves}, {Hassani}, {Henny}, {Henshaw}, {Hoyer},
  {Hughes}, {Jeffreson}, {Jimenez-Donaire}, {Kim}, {Kim}, {Klessen}, {Koch},
  {Kreckel}, {Kruijssen}, {Li}, {Liu}, {Lopez}, {Maschmann}, {Mayker Chen},
  {Meidt}, {Murphy}, {Neumann}, {Neumayer}, {Pan}, {Pessa}, {Pety},
  {Querejeta}, {Pinna}, {Jimena Rodr{\i}guez}, {Saito}, {Sanchez-Blazquez},
  {Santoro}, {Sardone}, {Smith}, {Sormani}, {Scheuermann}, {Stuber}, {Sutter},
  {Sun}, {Teng}, {Tress}, {Usero}, {Watkins}, {Whitmore}, \&
  {Razza}}]{LEE_PHANGSJWST}
{Lee}, J.~C., {Sandstrom}, K.~M., {Leroy}, A.~K., {et~al.} 2022{\natexlab{a}},
  arXiv e-prints, arXiv:2212.02667.
\newblock \doarXiv{2212.02667}

\bibitem[{{Lee} {et~al.}(2022{\natexlab{b}}){Lee}, {Whitmore}, {Thilker},
  {Deger}, {Larson}, {Ubeda}, {Anand}, {Boquien}, {Chandar}, {Dale},
  {Emsellem}, {Leroy}, {Rosolowsky}, {Schinnerer}, {Schmidt}, {Lilly},
  {Turner}, {Van Dyk}, {White}, {Barnes}, {Belfiore}, {Bigiel}, {Blanc}, {Cao},
  {Chevance}, {Congiu}, {Egorov}, {Glover}, {Grasha}, {Groves}, {Henshaw},
  {Hughes}, {Klessen}, {Koch}, {Kreckel}, {Kruijssen}, {Liu}, {Lopez},
  {Mayker}, {Meidt}, {Murphy}, {Pan}, {Pety}, {Querejeta}, {Razza}, {Saito},
  {S{\'a}nchez-Bl{\'a}zquez}, {Santoro}, {Sardone}, {Scheuermann}, {Schruba},
  {Sun}, {Usero}, {Watkins}, \& {Williams}}]{Lee2022}
{Lee}, J.~C., {Whitmore}, B.~C., {Thilker}, D.~A., {et~al.} 2022{\natexlab{b}},
  \apjs, 258, 10, \dodoi{10.3847/1538-4365/ac1fe5}

\bibitem[{{Lena} {et~al.}(2016){Lena}, {Robinson}, {Storchi-Bergmann}, {Couto},
  {Schnorr-M{\"u}ller}, \& {Riffel}}]{lena2016}
{Lena}, D., {Robinson}, A., {Storchi-Bergmann}, T., {et~al.} 2016, \mnras, 459,
  4485, \dodoi{10.1093/mnras/stw896}

\bibitem[{{Leroy} {et~al.}(subm.{\natexlab{a}})}]{LEROY2_PHANGSJWST}
{Leroy}, A., {et~al.} subm.{\natexlab{a}}, \apjl

\bibitem[{{Leroy} {et~al.}(subm.{\natexlab{b}})}]{LEROY1_PHANGSJWST}
---. subm.{\natexlab{b}}, \apjl

\bibitem[{{Leroy} {et~al.}(2021{\natexlab{a}}){Leroy}, {Schinnerer}, {Hughes},
  {Rosolowsky}, {Pety}, {Schruba}, {Usero}, {Blanc}, {Chevance}, {Emsellem},
  {Faesi}, {Herrera}, {Liu}, {Meidt}, {Querejeta}, {Saito}, {Sandstrom}, {Sun},
  {Williams}, {Anand}, {Barnes}, {Behrens}, {Belfiore}, {Benincasa},
  {Be{\v{s}}li{\'c}}, {Bigiel}, {Bolatto}, {den Brok}, {Cao}, {Chandar},
  {Chastenet}, {Chiang}, {Congiu}, {Dale}, {Deger}, {Eibensteiner}, {Egorov},
  {Garc{\'\i}a-Rodr{\'\i}guez}, {Glover}, {Grasha}, {Henshaw}, {Ho}, {Kepley},
  {Kim}, {Klessen}, {Kreckel}, {Koch}, {Kruijssen}, {Larson}, {Lee}, {Lopez},
  {Machado}, {Mayker}, {McElroy}, {Murphy}, {Ostriker}, {Pan}, {Pessa},
  {Puschnig}, {Razza}, {S{\'a}nchez-Bl{\'a}zquez}, {Santoro}, {Sardone},
  {Scheuermann}, {Sliwa}, {Sormani}, {Stuber}, {Thilker}, {Turner}, {Utomo},
  {Watkins}, \& {Whitmore}}]{Leroy2021-Survey}
{Leroy}, A.~K., {Schinnerer}, E., {Hughes}, A., {et~al.} 2021{\natexlab{a}},
  \apjs, 257, 43, \dodoi{10.3847/1538-4365/ac17f3}

\bibitem[{{Leroy} {et~al.}(2021{\natexlab{b}}){Leroy}, {Hughes}, {Liu}, {Pety},
  {Rosolowsky}, {Saito}, {Schinnerer}, {Schruba}, {Usero}, {Faesi}, {Herrera},
  {Chevance}, {Hygate}, {Kepley}, {Koch}, {Querejeta}, {Sliwa}, {Will},
  {Wilson}, {Anand}, {Barnes}, {Belfiore}, {Be{\v{s}}li{\'c}}, {Bigiel},
  {Blanc}, {Bolatto}, {Boquien}, {Cao}, {Chandar}, {Chastenet}, {Chiang},
  {Congiu}, {Dale}, {Deger}, {den Brok}, {Eibensteiner}, {Emsellem},
  {Garc{\'\i}a-Rodr{\'\i}guez}, {Glover}, {Grasha}, {Groves}, {Henshaw},
  {Jim{\'e}nez Donaire}, {Kim}, {Klessen}, {Kreckel}, {Kruijssen}, {Larson},
  {Lee}, {Mayker}, {McElroy}, {Meidt}, {Mok}, {Pan}, {Puschnig}, {Razza},
  {S{\'a}nchez-Bl'azquez}, {Sandstrom}, {Santoro}, {Sardone}, {Scheuermann},
  {Sun}, {Thilker}, {Turner}, {Ubeda}, {Utomo}, {Watkins}, \&
  {Williams}}]{Leroy2021-Pipeline}
{Leroy}, A.~K., {Hughes}, A., {Liu}, D., {et~al.} 2021{\natexlab{b}}, \apjs,
  255, 19, \dodoi{10.3847/1538-4365/abec80}

\bibitem[{{Levy} {et~al.}(2021){Levy}, {Bolatto}, {Leroy}, {Emig}, {Gorski},
  {Krieger}, {Lenki{\'c}}, {Meier}, {Mills}, {Ott}, {Rosolowsky}, {Tarantino},
  {Veilleux}, {Walter}, {Wei{\ss}}, \& {Zwaan}}]{Levy2021}
{Levy}, R.~C., {Bolatto}, A.~D., {Leroy}, A.~K., {et~al.} 2021, \apj, 912, 4,
  \dodoi{10.3847/1538-4357/abec84}

\bibitem[{{Li} \& {Zhang}(2020)}]{li2020}
{Li}, G.-X., \& {Zhang}, C.-P. 2020, \apj, 897, 89,
  \dodoi{10.3847/1538-4357/ab8c47}

\bibitem[{{Lindblad} {et~al.}(1996){Lindblad}, {Lindblad}, \&
  {Athanassoula}}]{Lindblad1996}
{Lindblad}, P.~A.~B., {Lindblad}, P.~O., \& {Athanassoula}, E. 1996, \aap, 313,
  65

\bibitem[{{Lindblad}(1999)}]{Lindblad1999}
{Lindblad}, P.~O. 1999, \aapr, 9, 221, \dodoi{10.1007/s001590050018}

\bibitem[{{Liu} {et~al.}(subm.)}]{LIU_PHANGSJWST}
{Liu}, D., {et~al.} subm., \apjl

\bibitem[{{Liu} {et~al.}(2021){Liu}, {Bureau}, {Blitz}, {Davis}, {Onishi},
  {Smith}, {North}, \& {Iguchi}}]{liu2021}
{Liu}, L., {Bureau}, M., {Blitz}, L., {et~al.} 2021, \mnras, 505, 4048,
  \dodoi{10.1093/mnras/stab1537}

\bibitem[{{Loni} {et~al.}(2021){Loni}, {Serra}, {Kleiner}, {Cortese},
  {Catinella}, {Koribalski}, {Jarrett}, {Molnar}, {Davis}, {Iodice},
  {Lee-Waddell}, {Loi}, {Maccagni}, {Peletier}, {Popping}, {Ramatsoku},
  {Smith}, \& {Zabel}}]{loni2021}
{Loni}, A., {Serra}, P., {Kleiner}, D., {et~al.} 2021, \aap, 648, A31,
  \dodoi{10.1051/0004-6361/202039803}

\bibitem[{{Loose} {et~al.}(1982){Loose}, {Kruegel}, \& {Tutukov}}]{Loose1982}
{Loose}, H.~H., {Kruegel}, E., \& {Tutukov}, A. 1982, \aap, 105, 342

\bibitem[{{Mandowara} {et~al.}(2022){Mandowara}, {Sormani}, {Sobacchi}, \&
  {Klessen}}]{mandowara2022}
{Mandowara}, Y., {Sormani}, M.~C., {Sobacchi}, E., \& {Klessen}, R.~S. 2022,
  \mnras, 513, 5052, \dodoi{10.1093/mnras/stac1214}

\bibitem[{{Martin}(1995)}]{Martin1995}
{Martin}, P. 1995, \aj, 109, 2428, \dodoi{10.1086/117461}

\bibitem[{{Mart{\'\i}n} {et~al.}(2021){Mart{\'\i}n}, {Mangum}, {Harada},
  {Costagliola}, {Sakamoto}, {Muller}, {Aladro}, {Tanaka}, {Yoshimura},
  {Nakanishi}, {Herrero-Illana}, {M{\"u}hle}, {Aalto}, {Behrens}, {Colzi},
  {Emig}, {Fuller}, {Garc{\'\i}a-Burillo}, {Greve}, {Henkel}, {Holdship},
  {Humire}, {Hunt}, {Izumi}, {Kohno}, {K{\"o}nig}, {Meier}, {Nakajima},
  {Nishimura}, {Padovani}, {Rivilla}, {Takano}, {van der Werf}, {Viti}, \&
  {Yan}}]{Martin2021}
{Mart{\'\i}n}, S., {Mangum}, J.~G., {Harada}, N., {et~al.} 2021, \aap, 656,
  A46, \dodoi{10.1051/0004-6361/202141567}

\bibitem[{{McMullin} {et~al.}(2007){McMullin}, {Waters}, {Schiebel}, {Young},
  \& {Golap}}]{2007ASPC..376..127M}
{McMullin}, J.~P., {Waters}, B., {Schiebel}, D., {Young}, W., \& {Golap}, K.
  2007, in Astronomical Society of the Pacific Conference Series, Vol. 376,
  Astronomical Data Analysis Software and Systems XVI, ed. R.~A. {Shaw},
  F.~{Hill}, \& D.~J. {Bell}, 127

\bibitem[{{Meidt} {et~al.}(2018){Meidt}, {Leroy}, {Rosolowsky}, {Kruijssen},
  {Schinnerer}, {Schruba}, {Pety}, {Blanc}, {Bigiel}, {Chevance}, {Hughes},
  {Querejeta}, \& {Usero}}]{meidt2018}
{Meidt}, S.~E., {Leroy}, A.~K., {Rosolowsky}, E., {et~al.} 2018, \apj, 854,
  100, \dodoi{10.3847/1538-4357/aaa290}

\bibitem[{{Moon} {et~al.}(2021){Moon}, {Kim}, {Kim}, \& {Ostriker}}]{moon2021}
{Moon}, S., {Kim}, W.-T., {Kim}, C.-G., \& {Ostriker}, E.~C. 2021, \apj, 914,
  9, \dodoi{10.3847/1538-4357/abfa93}

\bibitem[{{Moon} {et~al.}(2022){Moon}, {Kim}, {Kim}, \& {Ostriker}}]{moon2022}
---. 2022, \apj, 925, 99, \dodoi{10.3847/1538-4357/ac3a7b}

\bibitem[{{Morganti} {et~al.}(1999){Morganti}, {Tsvetanov}, {Gallimore}, \&
  {Allen}}]{Morganti1999}
{Morganti}, R., {Tsvetanov}, Z.~I., {Gallimore}, J., \& {Allen}, M.~G. 1999,
  \aaps, 137, 457, \dodoi{10.1051/aas:1999258}

\bibitem[{{Morokuma-Matsui} {et~al.}(2022){Morokuma-Matsui}, {Bekki}, {Wang},
  {Serra}, {Koyama}, {Morokuma}, {Egusa}, {For}, {Nakanishi}, {Koribalski},
  {Okamoto}, {Kodama}, {Lee}, {Maccagni}, {Miura}, {Espada}, {Takeuchi},
  {Yang}, {Lee}, {Ueda}, \& {Matsushita}}]{Morokuma-Matsui2022}
{Morokuma-Matsui}, K., {Bekki}, K., {Wang}, J., {et~al.} 2022, arXiv e-prints,
  arXiv:2210.08699.
\newblock \doarXiv{2210.08699}

\bibitem[{{Morris} \& {Serabyn}(1996)}]{Morris1996}
{Morris}, M., \& {Serabyn}, E. 1996, \araa, 34, 645,
  \dodoi{10.1146/annurev.astro.34.1.645}

\bibitem[{{Pan} {et~al.}(2022){Pan}, {Schinnerer}, {Hughes}, {Leroy}, {Groves},
  {Barnes}, {Belfiore}, {Bigiel}, {Blanc}, {Cao}, {Chevance}, {Congiu}, {Dale},
  {Eibensteiner}, {Emsellem}, {Faesi}, {Glover}, {Grasha}, {Herrera}, {Ho},
  {Klessen}, {Kruijssen}, {Lang}, {Liu}, {McElroy}, {Meidt}, {Murphy}, {Pety},
  {Querejeta}, {Razza}, {Rosolowsky}, {Saito}, {Santoro}, {Schruba}, {Sun},
  {Tomi{\v{c}}i{\'c}}, {Usero}, {Utomo}, \& {Williams}}]{pan2022}
{Pan}, H.-A., {Schinnerer}, E., {Hughes}, A., {et~al.} 2022, \apj, 927, 9,
  \dodoi{10.3847/1538-4357/ac474f}

\bibitem[{{Querejeta} {et~al.}(2016){Querejeta}, {Schinnerer},
  {Garc{\'\i}a-Burillo}, {Bigiel}, {Blanc}, {Colombo}, {Hughes}, {Kreckel},
  {Leroy}, {Meidt}, {Meier}, {Pety}, \& {Sliwa}}]{querejeta2016}
{Querejeta}, M., {Schinnerer}, E., {Garc{\'\i}a-Burillo}, S., {et~al.} 2016,
  \aap, 593, A118, \dodoi{10.1051/0004-6361/201628674}

\bibitem[{{Renaud} {et~al.}(2021){Renaud}, {Romeo}, \& {Agertz}}]{renaud+2021}
{Renaud}, F., {Romeo}, A.~B., \& {Agertz}, O. 2021, \mnras, 508, 352,
  \dodoi{10.1093/mnras/stab2604}

\bibitem[{{Renaud} {et~al.}(2013){Renaud}, {Bournaud}, {Emsellem}, {Elmegreen},
  {Teyssier}, {Alves}, {Chapon}, {Combes}, {Dekel}, {Gabor}, {Hennebelle}, \&
  {Kraljic}}]{renaud+2013}
{Renaud}, F., {Bournaud}, F., {Emsellem}, E., {et~al.} 2013, \mnras, 436, 1836,
  \dodoi{10.1093/mnras/stt1698}

\bibitem[{{Rigby} {et~al.}(2022){Rigby}, {Perrin}, {McElwain}, {Kimble},
  {Friedman}, {Lallo}, {Doyon}, {Feinberg}, {Ferruit}, {Glasse}, \&
  et~al.}]{Rigby2022}
{Rigby}, J., {Perrin}, M., {McElwain}, M., {et~al.} 2022, arXiv e-prints,
  arXiv:2207.05632.
\newblock \doarXiv{2207.05632}

\bibitem[{{Robitaille} \& {Bressert}(2012)}]{2012ascl.soft08017R}
{Robitaille}, T., \& {Bressert}, E. 2012, {APLpy: Astronomical Plotting Library
  in Python},  Astrophysics Source Code Library.
\newblock \doeprint{1208.017}

\bibitem[{{Sakamoto} {et~al.}(2007){Sakamoto}, {Ho}, {Mao}, {Matsushita}, \&
  {Peck}}]{Sakamoto2007}
{Sakamoto}, K., {Ho}, P. T.~P., {Mao}, R.-Q., {Matsushita}, S., \& {Peck},
  A.~B. 2007, \apj, 654, 782, \dodoi{10.1086/509775}

\bibitem[{{Sakamoto} {et~al.}(1999){Sakamoto}, {Okumura}, {Ishizuki}, \&
  {Scoville}}]{Sakamoto1999}
{Sakamoto}, K., {Okumura}, S.~K., {Ishizuki}, S., \& {Scoville}, N.~Z. 1999,
  \apj, 525, 691, \dodoi{10.1086/307910}

\bibitem[{{Sandqvist} {et~al.}(1995){Sandqvist}, {Joersaeter}, \&
  {Lindblad}}]{Sandqvist1995}
{Sandqvist}, A., {Joersaeter}, S., \& {Lindblad}, P.~O. 1995, \aap, 295, 585

\bibitem[{{Sandstrom} {et~al.}(subm.)}]{SANDSTROM1_PHANGSJWST}
{Sandstrom}, K., {et~al.} subm., \apjl

\bibitem[{{Schinnerer} {et~al.}(2019){Schinnerer}, {Hughes}, {Leroy}, {Groves},
  {Blanc}, {Kreckel}, {Bigiel}, {Chevance}, {Dale}, {Emsellem}, {Faesi},
  {Glover}, {Grasha}, {Henshaw}, {Hygate}, {Kruijssen}, {Meidt}, {Pety},
  {Querejeta}, {Rosolowsky}, {Saito}, {Schruba}, {Sun}, \&
  {Utomo}}]{schinnerer2019}
{Schinnerer}, E., {Hughes}, A., {Leroy}, A., {et~al.} 2019, \apj, 887, 49,
  \dodoi{10.3847/1538-4357/ab50c2}

\bibitem[{{Seo} \& {Kim}(2013)}]{Seo2013}
{Seo}, W.-Y., \& {Kim}, W.-T. 2013, \apj, 769, 100,
  \dodoi{10.1088/0004-637X/769/2/100}

\bibitem[{{Seo} {et~al.}(2019){Seo}, {Kim}, {Kwak}, {Hsieh}, {Han}, \&
  {Hopkins}}]{Seo2019}
{Seo}, W.-Y., {Kim}, W.-T., {Kwak}, S., {et~al.} 2019, \apj, 872, 5,
  \dodoi{10.3847/1538-4357/aafc5f}

\bibitem[{{Sheth} {et~al.}(2002){Sheth}, {Vogel}, {Regan}, {Teuben}, {Harris},
  \& {Thornley}}]{Sheth2002}
{Sheth}, K., {Vogel}, S.~N., {Regan}, M.~W., {et~al.} 2002, \aj, 124, 2581,
  \dodoi{10.1086/343835}

\bibitem[{{Sheth} {et~al.}(2005){Sheth}, {Vogel}, {Regan}, {Thornley}, \&
  {Teuben}}]{Sheth2005}
{Sheth}, K., {Vogel}, S.~N., {Regan}, M.~W., {Thornley}, M.~D., \& {Teuben},
  P.~J. 2005, \apj, 632, 217, \dodoi{10.1086/432409}

\bibitem[{{Sormani} \& {Barnes}(2019)}]{Sormani2019}
{Sormani}, M.~C., \& {Barnes}, A.~T. 2019, \mnras, 484, 1213,
  \dodoi{10.1093/mnras/stz046}

\bibitem[{{Sormani} {et~al.}(2015){Sormani}, {Binney}, \&
  {Magorrian}}]{Sormani2015}
{Sormani}, M.~C., {Binney}, J., \& {Magorrian}, J. 2015, \mnras, 449, 2421,
  \dodoi{10.1093/mnras/stv441}

\bibitem[{{Sormani} {et~al.}(2020){Sormani}, {Tress}, {Glover}, {Klessen},
  {Battersby}, {Clark}, {Hatchfield}, \& {Smith}}]{Sormani2020}
{Sormani}, M.~C., {Tress}, R.~G., {Glover}, S. C.~O., {et~al.} 2020, \mnras,
  497, 5024, \dodoi{10.1093/mnras/staa1999}

\bibitem[{{Sormani} {et~al.}(2018){Sormani}, {Tre{\ss}}, {Ridley}, {Glover},
  {Klessen}, {Binney}, {Magorrian}, \& {Smith}}]{Sormani2018}
{Sormani}, M.~C., {Tre{\ss}}, R.~G., {Ridley}, M., {et~al.} 2018, \mnras, 475,
  2383, \dodoi{10.1093/mnras/stx3258}

\bibitem[{{Spoon} {et~al.}(2007){Spoon}, {Marshall}, {Houck}, {Elitzur}, {Hao},
  {Armus}, {Brandl}, \& {Charmandaris}}]{spoon2007}
{Spoon}, H.~W.~W., {Marshall}, J.~A., {Houck}, J.~R., {et~al.} 2007, \apjl,
  654, L49, \dodoi{10.1086/511268}

\bibitem[{{Stark} {et~al.}(2004){Stark}, {Martin}, {Walsh}, {Xiao}, {Lane}, \&
  {Walker}}]{Stark2004}
{Stark}, A.~A., {Martin}, C.~L., {Walsh}, W.~M., {et~al.} 2004, \apjl, 614,
  L41, \dodoi{10.1086/425304}

\bibitem[{{Stevens} {et~al.}(1999){Stevens}, {Forbes}, \&
  {Norris}}]{Stevens1999}
{Stevens}, I.~R., {Forbes}, D.~A., \& {Norris}, R.~P. 1999, \mnras, 306, 479,
  \dodoi{10.1046/j.1365-8711.1999.02543.x}

\bibitem[{{Storchi-Bergmann} \& {Bonatto}(1991)}]{Storchi-Bergmann1991}
{Storchi-Bergmann}, T., \& {Bonatto}, C.~J. 1991, \mnras, 250, 138,
  \dodoi{10.1093/mnras/250.1.138}

\bibitem[{{Sun} {et~al.}(2022){Sun}, {Leroy}, {Rosolowsky}, {Hughes},
  {Schinnerer}, {Schruba}, {Koch}, {Blanc}, {Chiang}, {Groves}, {Liu}, {Meidt},
  {Pan}, {Pety}, {Querejeta}, {Saito}, {Sandstrom}, {Sardone}, {Usero},
  {Utomo}, {Williams}, {Barnes}, {Benincasa}, {Bigiel}, {Bolatto}, {Boquien},
  {Chevance}, {Dale}, {Deger}, {Emsellem}, {Glover}, {Grasha}, {Henshaw},
  {Klessen}, {Kreckel}, {Kruijssen}, {Ostriker}, \& {Thilker}}]{sun2022}
{Sun}, J., {Leroy}, A.~K., {Rosolowsky}, E., {et~al.} 2022, \aj, 164, 43,
  \dodoi{10.3847/1538-3881/ac74bd}

\bibitem[{{Tan}(2000)}]{tan2000}
{Tan}, J.~C. 2000, \apj, 536, 173, \dodoi{10.1086/308905}

\bibitem[{{Teng} {et~al.}(2022){Teng}, {Sandstrom}, {Sun}, {Leroy}, {Johnson},
  {Bolatto}, {Kruijssen}, {Schruba}, {Usero}, {Barnes}, {Bigiel}, {Blanc},
  {Groves}, {Israel}, {Liu}, {Rosolowsky}, {Schinnerer}, {Smith}, \&
  {Walter}}]{Teng2022}
{Teng}, Y.-H., {Sandstrom}, K.~M., {Sun}, J., {et~al.} 2022, \apj, 925, 72,
  \dodoi{10.3847/1538-4357/ac382f}

\bibitem[{{Teyssier}(2002)}]{teyssier2002}
{Teyssier}, R. 2002, \aap, 385, 337, \dodoi{10.1051/0004-6361:20011817}

\bibitem[{{Toomre}(1964)}]{toomre1964}
{Toomre}, A. 1964, \apj, 139, 1217, \dodoi{10.1086/147861}

\bibitem[{{Tress} {et~al.}(2020){Tress}, {Sormani}, {Glover}, {Klessen},
  {Battersby}, {Clark}, {Hatchfield}, \& {Smith}}]{tress2020}
{Tress}, R.~G., {Sormani}, M.~C., {Glover}, S. C.~O., {et~al.} 2020, \mnras,
  499, 4455, \dodoi{10.1093/mnras/staa3120}

\bibitem[{{Turk} {et~al.}(2011){Turk}, {Smith}, {Oishi}, {Skory}, {Skillman},
  {Abel}, \& {Norman}}]{Turk2011}
{Turk}, M.~J., {Smith}, B.~D., {Oishi}, J.~S., {et~al.} 2011, The Astrophysical
  Journal Supplement Series, 192, 9, \dodoi{10.1088/0067-0049/192/1/9}

\bibitem[{{Turner} {et~al.}(2021){Turner}, {Dale}, {Lee}, {Boquien}, {Chandar},
  {Deger}, {Larson}, {Mok}, {Thilker}, {Ubeda}, {Whitmore}, {Belfiore},
  {Bigiel}, {Blanc}, {Emsellem}, {Grasha}, {Groves}, {Klessen}, {Kreckel},
  {Kruijssen}, {Leroy}, {Rosolowsky}, {Sanchez-Blazquez}, {Schinnerer},
  {Schruba}, {Van Dyk}, \& {Williams}}]{turner2021}
{Turner}, J.~A., {Dale}, D.~A., {Lee}, J.~C., {et~al.} 2021, \mnras, 502, 1366,
  \dodoi{10.1093/mnras/stab055}

\bibitem[{{van den Bosch}(2016)}]{vandenBosch2016}
{van den Bosch}, R. C.~E. 2016, \apj, 831, 134,
  \dodoi{10.3847/0004-637X/831/2/134}

\bibitem[{{Veilleux} {et~al.}(2020){Veilleux}, {Maiolino}, {Bolatto}, \&
  {Aalto}}]{Veilleux2020}
{Veilleux}, S., {Maiolino}, R., {Bolatto}, A.~D., \& {Aalto}, S. 2020, \aapr,
  28, 2, \dodoi{10.1007/s00159-019-0121-9}

\bibitem[{{Veilleux} {et~al.}(2003){Veilleux}, {Shopbell}, {Rupke},
  {Bland-Hawthorn}, \& {Cecil}}]{Veilleux2003}
{Veilleux}, S., {Shopbell}, P.~L., {Rupke}, D.~S., {Bland-Hawthorn}, J., \&
  {Cecil}, G. 2003, \aj, 126, 2185, \dodoi{10.1086/379000}

\bibitem[{{Venturi} {et~al.}(2018){Venturi}, {Nardini}, {Marconi}, {Carniani},
  {Mingozzi}, {Cresci}, {Mannucci}, {Risaliti}, {Maiolino}, {Balmaverde},
  {Bongiorno}, {Brusa}, {Capetti}, {Cicone}, {Ciroi}, {Feruglio}, {Fiore},
  {Gallazzi}, {La Franca}, {Mainieri}, {Matsuoka}, {Nagao}, {Perna},
  {Piconcelli}, {Sani}, {Tozzi}, \& {Zibetti}}]{Venturi2018}
{Venturi}, G., {Nardini}, E., {Marconi}, A., {et~al.} 2018, \aap, 619, A74,
  \dodoi{10.1051/0004-6361/201833668}

\bibitem[{{Ward} {et~al.}(2022){Ward}, {Kruijssen}, {Chevance}, {Kim}, \&
  {Longmore}}]{ward2022}
{Ward}, J.~L., {Kruijssen}, J.~M.~D., {Chevance}, M., {Kim}, J., \& {Longmore},
  S.~N. 2022, \mnras, 516, 4025, \dodoi{10.1093/mnras/stac2467}

\bibitem[{{Watkins} {et~al.}(2022){Watkins}, {Barnes}, {Henny}, {Kim},
  {Kreckel}, {Meidt}, {Klessen}, {Glover}, {Williams}, {Keller}, {Leroy},
  {Rosolowsky}, {Boquien}, {Anand}, {Belfiore}, {Bigiel}, {Blanc}, {Cao},
  {Chandar}, {Mayker Chen}, {Chevance}, {Congiu}, {Dale}, {Deger}, {Egorov},
  {Emsellem}, {Faesi}, {Grasha}, {Groves}, {Hassani}, {Henshaw}, {Herrera},
  {Hughes}, {Jeffreson}, {Jimenez-Donaire}, {Koch}, {Kruijssen}, {Larson},
  {Liu}, {Lopez}, {Pessa}, {Pety}, {Querejeta}, {Saito}, {Sandstrom},
  {Scheuermann}, {Schinnerer}, {Sormani}, {Stuber}, {Thilker}, {Usero}, \&
  {Whitmore}}]{WATKINS_PHANGSJWST}
{Watkins}, E.~J., {Barnes}, A., {Henny}, K.~F., {et~al.} 2022, arXiv e-prints,
  arXiv:2212.00811.
\newblock \doarXiv{2212.00811}

\bibitem[{{Whitmore} {et~al.}(subm.)}]{WHITMORE_PHANGSJWST}
{Whitmore}, B., {et~al.} subm., \apjl

\end{thebibliography}

\end{document}